\documentclass[a4paper,twoside]{article}
\newcommand{\affil}[1]{$^{\rm #1}$}
\usepackage[authoryear]{natbib}
\bibpunct{(}{)}{;}{a}{}{,}
\usepackage{graphicx}
\usepackage{amssymb}
\usepackage{rotating}
\usepackage{psfrag}
\usepackage{color}
\date{} 
\newcommand{\arcsec}{$^{\prime\prime}$}

\newcommand{\tel}{ASKAP}
\newcommand{\sfind}{{\sc sfind}}
\newcommand{\imsad}{{\sc imsad}}

\newcommand{\tesla}{{\sc Aegean}}
\newcommand{\fragnames}{\psfrag{sex}{SE}\psfrag{paaadding}{\sfind}\psfrag{selavy}{Selavy}\psfrag{imsad}{\imsad}\psfrag{tesla}{\tesla}}

\newcommand{\changes}[1]{\textcolor{black}{#1}}

\title{\large\bf\flushleft VAST: An ASKAP Survey for Variables and Slow Transients}

\author{\parbox{\textwidth}{\flushleft
\vspace{-0.5cm}
{\it Tara Murphy\affil{A, B, C, V}, 
Shami Chatterjee\affil{D}, 
David L. Kaplan\affil{E},
Jay Banyer\affil{A, C},
Martin E. Bell\affil{A, C},
Hayley E. Bignall\affil{F}, 
Geoffrey C. Bower\affil{G},
Robert Cameron\affil{H},
David M. Coward\affil{I},
James M. Cordes\affil{D},
Steve Croft\affil{G},
James R. Curran\affil{B},
S. G. Djorgovski\affil{J},
Sean A. Farrell\affil{A, C},
Dale A. Frail\affil{K},
B. M. Gaensler\affil{A, C},
Duncan K. Galloway\affil{L},
Bruce Gendre\affil{M},
Anne J. Green\affil{A, C},
Paul J. Hancock\affil{A, C}, 
Simon Johnston\affil{N}, 
Atish Kamble\affil{E},
Casey J. Law\affil{G},
T. Joseph W. Lazio\affil{O},
Kitty K. Lo\affil{A, C},
Jean-Pierre Macquart\affil{F, C},
Nanda Rea\affil{P},
Umaa Rebbapragada\affil{O},
Cormac Reynolds\affil{F},
Stuart D. Ryder\affil{Q},
Brian Schmidt\affil{R, C},
Roberto Soria\affil{F},
Ingrid H. Stairs\affil{S},
Steven J. Tingay\affil{F, C},
Ulf Torkelsson\affil{T},
Kiri Wagstaff\affil{O},
Mark Walker\affil{U},
Randall B. Wayth\affil{F} 
Peter K. G. Williams\affil{G} } \\
\vspace{0.4cm}
{\small \affil{A}\,Sydney Institute for Astronomy, School of Physics, The University of Sydney, NSW 2006, Australia} \\
{\small \affil{B}\,School of Information Technologies, The University of Sydney, NSW 2006, Australia} \\
{\small \affil{C}\,ARC Centre of Excellence for All-sky Astrophysics (CAASTRO)} \\
{\small \affil{D}\,Department of Astronomy, Cornell University, Ithaca, NY 14853, USA} \\
{\small \affil{E}\,Physics Dept., University of Wisconsin --- Milwaukee, Milwaukee WI 53211, USA} \\
{\small \affil{F}\,ICRAR --- Curtin University, GPO Box U1987A, Perth, WA, Australia} \\
{\small \affil{G}\,Astronomy Department, University of California, Berkeley, Berkeley, CA 94720-3411, USA} \\
{\small \affil{H}\,Stanford University, Stanford, CA 94305, USA} \\
{\small \affil{I}\,School of Physics, University of Western Australia, Crawley WA 6009, Australia} \\
{\small \affil{J}\,California Institute of Technology, Pasadena, CA 91125, USA} \\
{\small \affil{K}\,National Radio Astronomy Observatory, P.O. Box O, Socorro, NM 87801, USA} \\
{\small \affil{L}\,Monash Centre for Astrophysics, School of Physics \& School of Mathematical Sciences, Monash University, VIC 3800, Australia} \\
{\small \affil{M}\,ASI Science Data Center, via Galileo Galilei, 00044 Frascati (RM), Italy} \\
{\small \affil{N}\,CSIRO Astronomy \& Space Science, Epping, NSW 1710, Australia} \\
{\small \affil{O}\,Jet Propulsion Laboratory, California Institute of Technology, Pasadena, CA 91109, USA} \\
{\small \affil{P}\,Institut de Ciencies de l'Espai (CSIC-IEEC), Campus UAB, Torre C5, 08193 Barcelona, Spain} \\
{\small \affil{Q}\,Australian Astronomical Observatory, Epping, NSW 1710, Australia} \\
{\small \affil{R}\,RSAA, Mount Stromlo Observatory, The Australian National University, ACT 2611, Australia} \\
{\small \affil{S}\,Dept. of Physics and Astronomy, University of British Columbia, 6224 Agricultural Road, Vancouver, BC V6T 1Z1 Canada} \\
{\small \affil{T}\,Department of Physics, University of Gothenburg, SE 412 96 Gothenburg, Sweden} \\
{\small \affil{U}\,Manly Astrophysics, 3/22 Cliff Street, Manly 2095, Australia} \\
{\small \affil{V}\,Corresponding author. Email: tara@physics.usyd.edu.au}}}

\begin{document}
\twocolumn[
\begin{changemargin}{.8cm}{.5cm}
\begin{minipage}{.9\textwidth}
\vspace{-1cm}
\maketitle
%
%
\small{\bf Abstract:}
The Australian Square Kilometre Array Pathfinder (ASKAP) will give us
an unprecedented opportunity to investigate the transient sky at radio
wavelengths.  In this paper we present VAST, an ASKAP survey for
Variables and Slow Transients.  VAST will exploit the wide-field
survey capabilities of ASKAP to enable the discovery and investigation
of variable and transient phenomena from the local to the
cosmological, including flare stars, intermittent pulsars, X-ray
binaries, magnetars, extreme scattering events, interstellar
scintillation, radio supernovae and orphan afterglows of gamma ray
bursts.  In addition, it will allow us to probe unexplored regions of
parameter space where new classes of transient sources may be
detected.  In this paper we review the known radio transient and
variable populations and the current results from blind radio
surveys. We outline a comprehensive program based on a multi-tiered
survey strategy to characterise the radio transient sky through
detection and monitoring of transient and variable sources on the
ASKAP imaging timescales of five seconds and greater.  We also present
an analysis of the expected source populations that we will be able to
detect with VAST.


\medskip
\medskip
\end{minipage}
\end{changemargin}
]
\small

\newcommand{\todo}{{\bf TODO:\ }}

\section{Introduction}
Many classes of astronomical objects are known to be 
variable radio sources, including the Sun, the planets, 
cool stars, stellar binary systems, neutron stars, supernovae, gamma-ray 
bursts and active galactic nuclei.
There have been extensive studies of these objects using targeted surveys,
but few blind, unbiased surveys for variable radio phenomena.
An effective survey for radio variability needs to maximise the metric
\begin{equation}
A \Omega \left(\frac{T}{\Delta t}\right)
\end{equation}
where $A$ is the effective collecting area of the telescope, $\Omega$ is the solid angle
coverage of the survey, 
$T$ is the total duration of the observations and $\Delta t$ is the time resolution \citep{cordes04,ska97}. 
Typically blind radio surveys have had either too small a field of
view and too little time coverage per field of view to allow a
comprehensive survey for transient and variable phenomena.

The Square Kilometre Array (SKA)\footnote{See http://www.skatelescope.org.} is a 
proposed future radio telescope \citep{dewdney09} that will be many times more
sensitive than any existing facility.
Should it go ahead, the SKA will be designed to explore five key science projects: galaxy
evolution, cosmology and dark energy; strong-field tests of gravity
using pulsars and black holes; the origin and evolution of cosmic
magnetism; probing the dark ages --- the first black holes and stars; and the cradle of life, searching
for life and planets.  In
addition, there is an awareness of building in (or not designing out)
the possibility of serendipitous discoveries.  In particular, the
dynamic radio sky is recognised as a rich and relatively unexplored
discovery space \citep{cordes04,ska97}.  Plans for the SKA are still
under development, with a Phase 1 instrument expected to be completed
in 2020, and a Phase 2 instrument in 2024.

The Australian Square Kilometre Array Pathfinder (ASKAP)\footnote{See
  http://www.atnf.csiro.au/projects/askap.} is a precursor and
technology development platform for the full SKA and will greatly
improve on our ability to detect radio transients and variable sources. 
Its moderately high sensitivity (RMS of $\sim$1~mJy~beam$^{-1}$ in
10~seconds), combined with a wide field of view (covering 30 square degrees
of sky in a single pointing) will enable fast, sensitive all-sky surveys \citep{johnston07,johnston08}.
ASKAP is currently under construction in Western Australia. The final
telescope will have 36 12-metre antennas, with a maximum baseline of
6~km.  The wide field of view is enabled by the
use of Phased Array Feeds \citep{chippendale10} of 100
dual-polarisation pixels that will operate over the range 700--1800
MHz.  A summary of the planned ASKAP specifications is given in
Table~\ref{t_specs}.
\begin{table}[t!]
\begin{center}
\caption{ASKAP Specifications\label{t_specs}}
\begin{tabular}{lr}
\hline
Number of dishes	& 36 \\
Dish diameter (m)	& 12 \\
Dish area (sq m)	& 113 \\
Total collecting area (sq m)	& 4072 \\
Aperture efficiency	& 0.8 \\
System temperature (K)	& 50 \\
Field-of-view (deg$^2$)	& 30 \\
Frequency range (MHz)	& 700--1800 \\
Bandwidth (MHz)	 & 300 \\
Maximum number of channels	& 16\,384 \\
Maximum baseline (km)	& 6 \\
RMS cont. sensitivity (10~s) & $640\mu$Jy/bm \\
RMS cont. sensitivity (1~hr) & $47\mu$Jy/bm \\
\hline
\end{tabular}
\end{center}
\end{table}
The 6-antenna Boolardy Engineering Test Array (BETA), is
expected to be completed in late 2012 and will allow some early
testing of the techniques required for the full ASKAP surveys.

ASKAP is primarily a survey instrument, and this is reflected in the time allocation policy.
In its initial five years of operation $75\%$ of the available
time will be scheduled for several major Survey Science Projects\footnote{The full list of projects
is available at http://www.atnf.csiro.au/projects/askap/ssps.html.} that were selected
through a competitive call for proposals \citep{ball09}.

In this paper we describe one of these projects, An ASKAP Survey for Variables and Slow Transients (VAST)\footnote{See http://www.askap.org/vast.}.
In the context of ASKAP, `slow' means variability or transient behaviour 
detectable on timescales greater than the cadence on which images will be performed (expected to
be 5--10~seconds).
A related project, The Commensal Real-Time ASKAP 
Fast-Transients Survey \citep[CRAFT;][]{macquart10}, will search for `fast' transients
on timescales shorter than 5~seconds. Different techniques are required to process and analyse data for fast
and slow transients, and from a scientific perspective this division roughly represents
a separation between coherent and incoherent emission processes.
The main science goals of VAST are:
\begin{itemize}
\item to detect and monitor `orphan' gamma-ray burst afterglows to understand their nature;
\item to conduct an unbiased survey of radio supernovae in the local Universe;
\item to determine the origin and nature of the structures responsible for extreme scattering events;
\item to make a direct detection of baryons in the intergalactic medium;
\item to discover flaring magnetars, intermittent or deeply nulling radio pulsars, and rotating radio transients through changes in their pulse-averaged emission;
\item to investigate intrinsic variability of active galactic nuclei;
\item to detect and monitor flare stars, cataclysmic variables and X-ray binaries in our Galaxy; and 
\item to discover previously unknown classes of objects.
\end{itemize}
To achieve these goals VAST will conduct three surveys, in addition to commensal observing with
other ASKAP projects: VAST-Wide, aimed at detecting rare, bright events, will survey $10\,000$ square degrees
per day to an RMS sensitivity of 0.5~mJy~beam$^{-1}$; VAST-Deep, aimed at detecting unknown source classes, will
survey $10\,000$ square degrees down to an RMS sensitivity of $50\mu$Jy~beam$^{-1}$; and VAST-Galactic which will cover
750 square degrees of the Galactic plane, plus the Large and Small Magellanic Clouds down to an RMS sensitivity of
0.1~mJy~beam$^{-1}$.
The full survey parameters are shown in Table~\ref{t_survey} and discussed in Section~\ref{s_survey}.
\begin{table*}[t!]
\begin{center}
\caption{Survey parameters for the VAST surveys (see Section~\ref{s_sspecs} for details). The two components of
the VAST-Deep survey are discussed further in the text. The Commensal column gives an indication of the 
amount of time that VAST will have access to through commensal observing with other ASKAP projects.}\label{t_survey}
\begin{tabular}{l|ccccc}
\hline
                      & VAST-Wide & \multicolumn{2}{c}{VAST-Deep} & VAST-Galactic & Commensal \\
 & & Multi-field & Single field & & \\
\hline
Observing time (hrs)  & 4380      & 3200 & 400   &   600   & 1.5 years  \\
Survey area (sq deg)  & 10\,000   & 10\,000 & 30  & 750 & 10\,000\\
Time per field        & 40~s & 1~hr & 1 hr & 16~min & 12 hours\\
Repeat                & daily & 7 times & daily & 64 times & none \\
Observing freq (MHz) & \multicolumn{5}{c}{1130--1430} \\
Bandwidth (MHz)     & \multicolumn{5}{c}{300} \\
RMS sensitivity (mJy~beam$^{-1}$) & 0.5 & \multicolumn{2}{c}{0.05} & 0.1 & 0.01\\ 
Field of view (sq deg) & \multicolumn{5}{c}{30} \\
Angular resolution & \multicolumn{5}{c}{10\arcsec} \\
Spectral resolution & \multicolumn{5}{c}{10~MHz} \\ 
Time resolution & \multicolumn{5}{c}{5~seconds} \\
Polarisation products & \multicolumn{5}{c}{IQUV} \\
\hline
\end{tabular}
\end{center}
\end{table*}

In Section~\ref{s_science} we motivate the VAST surveys by discussing the science goals in more
detail, and explain how they fit in the context of existing efforts. We also
make predictions about the likely source populations that will be
observed by VAST.
In Section~\ref{s_blind} we review
previous and current blind radio surveys, as well as archival
studies. 
In Section~\ref{s_survey} we describe the proposed VAST survey
strategy in detail. Finally, in Section~\ref{s_software} we discuss the
design of the VAST transient detection pipeline.

\section{VAST Science}\label{s_science}
Astronomical transient phenomena are diverse in nature, but they can
be broadly classified on the basis of their underlying physical
mechanism into one of four general categories: explosions, propagation
effects, accretion-driven, and magnetic field-driven events.
In this section we
discuss the main VAST science goals in each of these areas and discuss
the expected populations for different types of sources.

\subsection{Explosions}

Gamma-ray bursts (GRBs) and supernovae (SNe) are some of the most energetic 
explosions in the Universe. Although they have been studied extensively
there are a number of unresolved questions about their nature, 
including the existence of radio orphan
afterglows \citep{levinson02}, the beaming fraction of gamma-ray
bursts \citep{sari99,dalal02}, and the current rate of massive star
formation in the Universe as traced by the cosmological radio
supernova rate.  These questions can be addressed by surveys
that observe a large area of sky to good sensitivity, on a regular
basis.  The capacity of VAST to detect these rare objects is orders of
magnitude greater than previous blind surveys such as those of
\citet{galyam06} or \citet{bower07}.

\subsubsection{Gamma-Ray Bursts}
The afterglow emission from gamma-ray bursts is generated via particle
acceleration in the decelerating blast wave. At late times, the peak
of the afterglow spectrum is at radio energies, and thus radio studies
provide the most interesting probes for GRBs science. Based on
previous studies \citep{chandra12}, we expect that the GRBs seen by
ASKAP will be long-lasting but intrinsically fainter on average than
shorter wavelength afterglows. Given the sensitivity of ASKAP, this
means that those GRBs studied by VAST will, in general, be the rare
but important nearby ($\sim$1500~Mpc) events. In the relativistic expansion
phase (up to 100 days after the initial explosion), ASKAP monitoring
in conjunction with optical and X-ray facilities will allow us to
measure physical parameters such as the total kinetic energy of the
explosion, the density of the circumburst medium and the geometry of
the outflow. 

At late times, when the blast wave has decelerated to sub-relativistic
velocities, the jetted outflows expand and become more isotropic.
Under these circumstances, the the total (i.e.  calorimetric) energy
of GRB blast waves may be estimated - a key but difficult parameter to
measure \citep{frail05}. Instead of relying on rare, nearby events,
VAST may also obtain a robust beaming-independent estimates GRB
energies from a large sample of single-epoch flux density measurements
(and limits) of afterglows at late times \citep{shivvers11}.

As the relativistic beaming is totally suppressed at late time, radio
monitoring is the best solution to identify `orphan afterglows', i.e.,
those GRBs whose jet is beamed away from us \citep{rhoads97}. While
they would lack a gamma-ray trigger by definition, such systems could
be distinguished from other classes of transients through the
evolution and luminosity of their light curves, their location
towards galaxies that resemble other GRB hosts, and their
probable association with
supernovae \citep{soderberg10}. The discovery of a population of
orphan afterglows would allow us to constrain the beaming fraction of
GRBs -- a key quantity for understanding the true GRB event rate. It
would also open up an order of magnitude more objects which can be
scrutinised in detail. This may also give insights on the jet
structure of GRBs 
\changes{
and clarify the puzzling lack of jet breaks from 
the {\it Swift} satellite \citep{racusin09}.
}

\subsubsection{Supernovae}
While thermonuclear (Type Ia) supernovae have proven to be extremely useful 
in measuring the acceleration of the Universe, no Type Ia SN has ever 
been detected at radio wavelengths \citep{panagia06,hancock11,chomiuk12}, indicating 
a very low density of circumstellar material. In contrast, many core-collapse 
SNe (those of Type Ib, Ic, and the various Type II sub-classes) have been 
detected and monitored at radio wavelengths \citep{weiler02,ryder04,soderberg07}.
Radio observations replay the late phases of stellar evolution (in reverse), and reveal 
clues about the late-time evolution of the progenitor system.

With VAST we will conduct an unbiased census of core-collapse SNe (a 
still undetermined fraction of which emit at radio wavelengths),
allowing us to match radio detections against optically-discovered SNe.
We will also detect new radio SNe that may have gone undetected in the optical or 
infrared due to significant amounts of dust \citep{kankare08}. 
The prevalence of such
heavily-obscured SNe is virtually unconstrained, with hints
that they may constitute a nontrivial subpopulation. Recently, SN~2008cs
\citep{kankare08} was detected in the IR and confirmed in the radio bands, and 
SN~2008iz \citep{brunthaler09} was serendipitously discovered in radio observations
of M\,82.  SN~2008iz has remained undetected in the optical, IR, and X-ray bands
despite sensitive searches \citep{brunthaler10}.
Discovery of new SNe is valuable for tying down the current rate of massive star formation 
in the Universe and for understanding how and when SNe explode in star-bursting galaxies.
Radio emission from core-collapse SNe is detectable within days, and they reach their peak luminosity at 1.4~GHz 
as much as one year after the explosion, making them ideal targets to
detect and study with the sensitivity 
and cadence of VAST (see Section~\ref{s_survey}).

Beyond the large population of standard core-collapse SNe, there exist
sub-classes of objects that are of particular interest.  For instance, radio
observations of a recent type Ib/c supernova, SN2009bb, showed that it
was expanding at mildly relativistic velocities ($\beta \approx 0.8$)
powered by the central engine \citep{soderberg10}.  Such SNe
are usually associated with GRBs and are discovered by their intense
but short-lived gamma-ray energy emission. Yet, SN2009bb had no detected
gamma-ray counterpart. It has been suggested that the blast-waves
produced in such engine-driven SNe could be responsible for the
acceleration of ultra-high energy cosmic-rays \citep{chakraborti11}.
However, the present limits on the neutrino flux expected as a
by-product of cosmic-ray production in GRBs are at least a factor of
3.7 below predictions \citep{abbasi12}.  Radio surveys with the
capability of completing the sample of these explosions will improve
our understanding of particle acceleration processes and the diversity
of core-collapse events \citep[see][for predictions on the detection
rates of core-collapse radio SNe for a range of
SKA pathfinder instruments]{lien11}.

\subsubsection{Gravitational Wave Counterparts}\label{s_gw}
Gravitational wave (GW) events are expected to be compact and involve
significant energies---both GRBs and SNe are considered to be among
likely sources of both electromagnetic and GW emission.  In an
analogous manner, other classes of GW events may also generate
significant transient radio emission, especially if they occur in
dense environments. 
We discuss the particular case of mergers of
neutron star binaries in Section~\ref{s_nsm}. For all classes of GW
events, however, localisation will be a significant challenge, with search
areas expected to be of order 10--100~deg$^2$ for planned GW
detectors \citep{fairhurst11}. The large instantaneous sky coverage of
ASKAP makes it particularly valuable in the search for
electromagnetic counterparts to GW events, although there are
significant uncertainties in the requisite search sensitivities and
timescales. The search area estimates mentioned above are computed for
the Advanced LIGO \citep{harry10} and Advanced VIRGO \citep{accadia11}
detectors. These systems are planned to come online in the 2014--2015
timeframe, meaning that ASKAP will be well-positioned to make
discoveries in this parameter space.

\subsection{Propagation}\label{sec:prop}

{\bf Plasma Propagation Effects:}
At centimetre wavelengths any object with an angular size $\lesssim
1\,$mas is subject to the effects of interstellar scintillation (ISS),
caused by inhomogeneities in the ionised interstellar medium (ISM) of
our Galaxy.  
Refractive ISS (RISS) is manifested as slow intensity variations with
a wide correlation bandwidth that is caused by large-scale refracting
structures in the ISM.  Much faster variations that have narrow
frequency scales are associated with diffraction from small-scale
irregularities (DISS).
Observations of the intensity fluctuations caused by ISS
can be used to infer both the small-scale structure of the ISM and
the structure of a scintillating source on scales down to $\sim
10-200$~microarcseconds for compact extragalactic sources and $\sim
10$~nanoarcseconds for some pulsars.

The density inhomogeneities in the ionised ISM are typically
characterised as a power law \citep{ars95}.  However, a small number
of pulsars have displayed strong fringing events in their dynamic
spectra, in which the normal, random appearance of the dynamic
spectrum caused by DISS 
is modified by the appearance of quasi-periodic fringes. Both
pulsars and AGN have displayed extreme scattering events (ESEs) in
their light curves, in which their flux densities exhibit marked
decreases ($\approx 50$\%) bracketed by modest increases; and a 
number of AGN have exhibited intra-day variability (IDV), marked
by flux density variations on time scales of hours.  Fringing events
are interpreted most naturally as the beating of two distinct images
of the pulsar and both ESEs and IDV can be interpreted as due to
lensing events through a plasma.  All of these phenomena could occur
if there is a temporary increase in the amount of power on scales of
order $10^{13}$~cm along or close to the the line of sight, e.g., as
due to a discrete ``cloud'' of electrons drifting in front of the
pulsar or AGN \citep{cw86,cfl98}. In the case of \hbox{IDV}, it may
also be the case that transient, high brightness temperature
structures are required to be produced in the cores of some
\hbox{AGN}.

The characteristics of the discrete structures inferred to be
responsible for these phenomena are such (density $> 100$~cm${}^{-3}$,
temperature $> 10^3$~K) that they would be highly overpressured with
respect to the bulk properties of the \hbox{ISM} \citep{kh88}, and
would likely be transient structures.
While there is some evidence to suggest that a small fraction of the
volume of the ISM is at a high pressure \citep{jt01}, the distribution
of such highly overpressured structures is not known, although the
characteristics of IDV in some AGN suggest that some structures are
extremely close to the Sun ($\sim 1-10$~pc).  The existence of such
highly over-pressured structures in the local ISM is hard to reconcile
with our current understanding of interstellar turbulence
\citep[e.g.][]{fiedler87b,walker07}.  Alternatively, these
phenomena may indicate that a simple power law is not a good
description of the spectrum of density inhomogeneities, which would
then constrain the nature of the plasma processes involved in the
generation and maintenance of the density inhomogeneities.

Because these propagation events are transient in nature, many of the
previous observations have been ``accidental,'' happening to catch a
source displaying such an event during the course of observations
designed for other reasons.  There have been relatively few surveys
designed specifically to try to find propagation events, with notable
examples being the Green Bank Interferometer (GBI) monitoring program
\citep{fdjws94,lwgffj01} and the MASIV survey
\citep{lovell03,lovell08}.  Moreover, because of either limited
sensitivity, limited field of view, or both, most surveys have only
been able to monitor a relatively small number ($\sim 100$) of
relatively strong sources ($\gtrsim 100$~mJy).

A wide-area, high cadence, sensitive survey would be able to place
strong constraints on both the spatial distribution of the structures
responsible for propagation events and the likelihood of a source
exhibiting such an event.  In turn, these much stronger constraints
can be used to infer properties of the ionised ISM and the
mechanisms by which such structures are generated.  As an illustration
of the kind of monitoring program that ASKAP will enable, a survey
could be designed to produce daily flux density measurements of
sources stronger than about 10~mJy with the capability to detect 50\%
flux density changes at the 10$\sigma$ level.  At this sensitivity
level, a survey of even only 1000~deg${}^2$ would still be able to
monitor nearly compact 50,000 sources.

\noindent{\bf Gravitational Lensing:}
Aside from the plasma propagation effects discussed in more detail
below, we also note that strong gravitational lensing of active 
galaxies by galaxy clusters can lead to an achromatic time-domain 
signal that contains information
about both the active galaxy and about the intervening space-time.
Radio observations of strong lensing have been conducted successfully
with existing facilities and those data have been used to constrain
cosmological parameters \citep[e.g][]{fassnacht99,myers03}.  These
operate by spatially separating the lensed components of the active
galaxy.  In general ASKAP will not have the angular resolution to 
observe such multiple imaging, but the large number of sources
monitored with regular cadence means that auto-correlation analysis
\citep[as performed by][]{barnacka11} can itself constrain lens time
delays.

\subsubsection{Pulse Broadening from Scattering}

The well known temporal broadening of pulsar pulses caused by
scattering from electron-density variations will affect any source of
short duration bursts, including those that will be targeted by VAST.
Pulse broadening from sources in the Galactic center can be broader
than 1000 sec at 1 GHz.  Bursts from extragalactic sources at
cosmological distances can similarly be broadened, especially if the
line of sight intersects another galaxy but also from the
intergalactic medium (IGM).  Scattering in the Milky Way, in other galaxies,
and in the IGM therefore define pulse-broadening horizons beyond which
pulses with specific widths will not be seen \changes{\citep[see,
  e.g.,][]{cordes12}}.  Pulse broadening from scattering can be
distinguished from source-intrinsic effects by identifying across the
ASKAP band the strong frequency dependence $\sim \propto \nu^{-4}$ of
scatter broadening.

\subsubsection{Extreme Scattering Events}\label{s_ese}
\begin{figure}[t!]
\begin{center}
\includegraphics[width=\columnwidth]{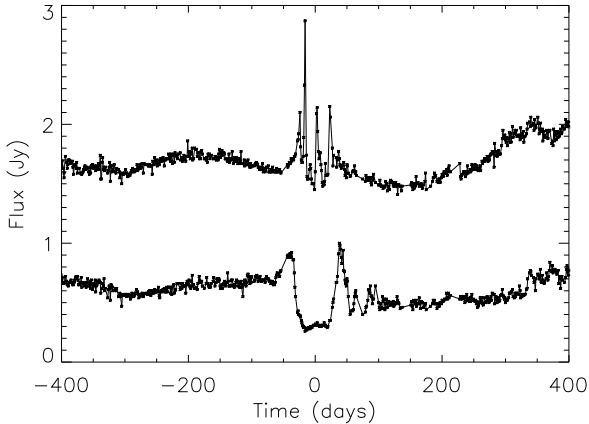}\vspace*{-2mm}%
\caption{\small%
An Extreme Scattering Event in Q0954$+$658 at 2.7~GHz (lower) and
8.1~GHz (upper) adapted from \citet{fiedler87b}.  For clarity, an
offset of 1~Jy has been added to the top trace.  The strong frequency
dependence of ESEs and the necessity of regular sampling of the light
curves over a long period is evident.  At ASKAP frequencies ($\sim
1$~GHz), the amplitude of the flux density decrease would likely have
been even larger, and the flux density would likely have increased to
an even higher value during the start of the event.
\label{f_ese}}
\end{center}
\end{figure}

Extreme scattering events (ESEs) are flux density
variations marked by decreases in flux density bracketed by increases
(Figure~\ref{f_ese}).  There is widespread agreement that ESEs are
caused by refractive lens-like density perturbations along the
line-of-sight to a background source \citep[e.g.][]{fiedler87b,romani87}.  The
refracting lenses which cause ESEs must be Galactic in origin and
probably lie within a few kpc, but in the 24 years since their
discovery, no satisfactory physical model has emerged for the ESE
phenomenon.  Analysis of existing data suggests that ESEs are unlike
any other identified component of the ISM and that they are present in
vast numbers in the Galaxy (with a density $\sim 10^3\,$pc$^{-3}$).
If the refracting properties are taken at face value, they imply
regions with pressures $\sim 1000$ times higher than the diffuse
ionized ISM. One suggestion is that they are associated with neutral,
self-gravitating gas clouds \citep{walker07}, in which case they would
make a substantial contribution to the total baryonic mass content of
the Galaxy.  Alternatively, they may represent small-scale density
enhancements created by converging flows in a turbulent medium
\citep{jt01}, but it is unclear whether such structures could persist
for the observed ESE durations.

A wide-area survey, with daily cadence would allow us to address the
physical composition and origin of the objects responsible for ESEs in
a comprehensive manner. \\

\noindent {\bf Population statistics:}
One of the more comprehensive ESE surveys \citep{fiedler94} provided a
total coverage of approximately 600~source-years; VAST has the
potential to achieve a coverage that is larger by a factor of roughly
300, searching for ESEs toward sources with much lower flux densities.
In turn, increased coverage will constrain the spatial density and
distribution of refracting lenses in the Galaxy, including whether
they are associated with particular Galactic structures
\citep{fpjd94}.  A larger number of ESEs will also allow investigation
into any potential diversity in light curve shapes, and whether any
such diversity can be understood in terms of simple models
\citep{cfl98} or the extent to which more complex models of the
density structures, background sources, or both are required. \\

\noindent {\bf Real-time characterisation of ESEs:} 
A straightforward approach to elucidating the nature of the structures
responsible for ESEs would be to conduct additional observations of a
source while it is undergoing an \hbox{ESE}.  Events like those seen
towards quasars B0954$+$658 and B1749$+$096 \citep{fiedler94} or
B1741$-$038 \citep{fjwg92,cff96} should be seen at the rate of roughly
60 per year in an all-sky monitoring survey with $\sim$1~mJy RMS
sensitivity, with $\sim$10 in progress at any given moment.  Thus, in
contrast to the GBI monitoring program, for which only one ESE was
identified while it was on-going, VAST will enable real-time
follow-up.

Some key follow-up measurements could be made at higher
frequencies, particularly in the optical to X-ray bands, which should
directly reveal the dense neutral gas clouds responsible for ESEs via
scattering, absorption and refraction \citep{draine98}.
Additionally, H\,\textsc{i} observations would reveal any associated
neutral structures, Faraday rotation measurements may constrain the
magnetic field strength within the structures, and high resolution
observations could confirm predicted changes in the apparent structure
of the source.  While there have been previous efforts in conducting
such observations \citep{cff96,lfd+00,lgcffj01}, only one source has
been observed in such a manner (B1741$-$038), and, for some of the
measurements, only a single observation during the actual ESE could be
obtained.  An all-sky survey could detect multiple ESEs during the
course of a year, allowing ample opportunity to trigger follow-up
observations with other telescopes and at other wavelengths.

\subsubsection{Interstellar Scintillation of AGN}\label{s_sagn}
Observations of decimetre variability in extragalactic sources
\citep[e.g.,][]{hunstead72} led quickly to the recognition that 
the sources needed to have extremely
compact components, be affected by interstellar scintillation, or both
\citep{ash77,sfgp89}.  The need for extremely compact components,
apparently violating the so-called inverse Compton catastrophe, became
even more acute with the discovery of ``flickering'' \citep{h84}, and
later intra-day variability.

The MASIV 5GHz VLA Survey \citep{lovell03,lovell08} and other
observations \citep{kc97,dennettthorpe02,bignall03,bignall06} have
established that the fast---intra-day and inter-day---flux density
variations at centimetre wavelengths exhibited by a subset of compact
quasars are predominantly caused by ISS rather than intrinsic
variability.  Intra-day variability refers to sources whose
characteristic timescale is less than 24~hours, while inter-day
variability refers to sources whose timescale exceeds this.

Scintillating AGN are of astrophysical interest because the small angular sizes
that they must possess in order to exhibit rapid ISS require
brightness temperatures near, or possibly in excess of, the
$10^{12}\,$K inverse Compton limit for incoherent synchrotron
radiation.  In several cases over the past decade, it appeared that
the Doppler boosting factors required in order to reconcile the
apparent brightness temperature with the inverse Compton limit were
well in excess of the typical values determined from Very Long
Baseline interferometric observations of quasars.  
Estimates of the Doppler boosting factors are subject to considerable
uncertainty related to the distance of the scattering material from
Earth, but many of the most rapid IDVs appear to require Doppler
factors of a few tens
\citep{macquart00,rickett02,bignall06,macquart07}.

More recently, it appears that the fastest manifestations of IDV are
associated with sources whose lines of sight intersect more local,
albeit much more inhomogeneous, patches of turbulence than previously
supposed.  Thus there has been a transfer of our ignorance away from
the physics of the AGN and onto the physics of interstellar
turbulence.  However, this issue is not yet conclusively resolved, as
little is known about the origin (or physical plausibility) of the
local, highly inhomogeneous scattering screens that need to be 
invoked to explain the observed scintillation properties of these
sources. 

The four-epoch MASIV survey \citep{lovell03,lovell08} detected
variability in 56\% of the 482 sources observed but the
characteristics of the variability changed from epoch to epoch.
While able to illustrate this ISS intermittency, the MASIV survey
itself, because
of its limited duration and cadence, could not address many of the
resulting questions, notably the extent to which the intermittency
results from changes in the ISM properties or changes in the sources.
In order to address the cause of the intermittency, a survey must have
the following characteristics.  It must observe a large number of
sources widely distributed on the sky so as to have robust
statistics and to be able to search for correlations expected
for ISS (e.g., a Galactic latitude dependence or with nearby
structures in the ISM).  The survey must measure both the total and
linear polarization intensity of the sources---ISM-driven
intermittency should result in the linear polarization scintillation
tracking that of the total intensity, while source-driven
intermittency should result in the linear polarization and total
intensity scintillating nearly independently of each other.  
The survey must be conducted with close to daily cadence to capture
the scintillation variations, and over a multi-year time scale in
order to separate changes in source structure or scattering medium
from changes associated with the annual cycle resulting from the
Earth's orbital velocity.

In addition to determining the cause of ISS intermittency and
mapping the scattering properties of the local ISM, such a survey
would address the existence of AGN emission above the inverse
Compton brightness limit and potentially constrain
the baryonic content of the IGM as a function of redshift.\\

\noindent {\bf The cause of ISS intermittency:}
By continuously observing a large sample of sources over a long
period, VAST will comprehensively determine the intermittency
properties of scintillating \hbox{AGN}. It will discriminate between
ISM-related and source-driven effects via polarization variations, as
discussed above.
A complication arises because annual cycles in the variability
timescale (caused by the Earth's velocity vector changing with respect
to the scattering screens) need to be distinguished from true IDV
intermittency.  However, these two phenomena can be separated by
making regular (daily) observations over the course of at least two
years.  The variations should be dissimilar between epochs separated
by a year if the scintillation is intermittent.  Intermittency, if
caused by the \hbox{ISM}, points to the physical nature of the
turbulent patches responsible for \hbox{IDV}. \\

\noindent {\bf AGN emission above the inverse Compton brightness limit:} 
In some cases fast scintillation may be attributable to the
appearance of micro-arcsecond structure within a source, associated
with an outburst \citep{macquart07}.  The rise and fade times of the
manifestation of fast variations allow us to determine the longevity
of micro-arcsecond components associated with \hbox{IDV}: The rate at
which the AGN supplies energy to power the bright emission that causes
the variability, and the rate at which energy losses cause its
eventual decay.  These rates can then be compared with specific AGN
processes to identify the mechanism associated with the super-Compton
radio emission. \\

\noindent {\bf Baryonic content of the intergalactic medium:} 
The MASIV survey showed that few sources above $z\sim2$ are variable
\citep{lovell08}.  If this apparent suppression of ISS as a function
of source redshift is confirmed by \hbox{VAST}, there are a number of
possible explanations: (1)~intrinsic source evolution;
(2)~gravitational microlensing; or~(3)~scattering in the turbulent
ionised IGM.  If the last is at least partially responsible, 
then the angular broadening
is caused by the cumulative effect of all the baryons in the ionised
\hbox{IGM}.
The sheer number of sources VAST can monitor will enable
us to probe in detail the evolution of structure in the ionised IGM as
a function of redshift.  

One potential complication with scintillation studies is that
the largest amplitude IDV is typically observed at frequencies
near~5~GHz.  At ASKAP frequencies, it is possible that the strength of
scattering, even toward high Galactic latitudes, and intrinsic source
diameters will result in large-amplitude scintillations being quenched
\citep{dc81}, although lower amplitude refractive scintillations
may still be exhibited \citep[e.g.,][]{dfj+87}.

\subsection{Accretion and Magnetism}

Along with explosions and propagation effects, accretion and magnetic
fields are frequently implicated in transient events.  Accretion onto
a compact object is a powerful energy source that drives both local
sources such as X-ray binaries (XRBs) and microquasars as well as
extragalactic blazars and AGN at cosmological
distances.  The accretion flow is often connected to an outflow in the
shape of a jet, which is the source of the radio emission.  The
intimate link between the accretion disc and the jet is demonstrated
by the fundamental plane of black hole activity that describes a
correlation between the mass of the black hole, the jet power, and the
accretion power \citep{mer2003a,fal2004a} in active galactic nuclei.
This relation can be extended to lighter black holes where it matches
the correlation between the X-ray and radio luminosities in XRB black
hole candidates (BHCs) that was found by \citet{gal2003a}.  Radio
emission is also observed from XRBs containing weakly magnetised neutron
stars, such as Z and atoll sources, although they are at least one order
of magnitude fainter in radio than the BHCs (e.g. \citealt{fender06}),
and a weak radio jet has also been observed from the cataclysmic
variable SS Cyg during a dwarf nova outburst \citep{kording08}.  There
is still not a good theoretical understanding of how accretion discs produce
outflows, but most models assume that the jet is
accelerated by a magnetic field \citep[e.g.][]{bla1982a}.  VAST will
allow us to characterise the full range of transient jet phenomena
across the fundamental plane.

Magnetic fields can also act in other ways.  They can convert the
rotational energy of their source into electromagnetic energy and the
mechanical energy of an outflow, as happens for instance in a radio pulsar,
or they can release their own energy in the form of flares on a star
or the radio emission from Jupiter and (possibly) Jupiter-like
extrasolar planets.  With VAST we will be able to reveal the unifying
physical principles that underlie such diverse behaviour.

\subsubsection{Intrinsic Variability in Active Galactic Nuclei}
As discussed in Section~\ref{s_sagn}, \citet{fiedler87a} observed a 
sample of 33 AGN, with flux densities
of a few Janskys, every day for more than 6 years. All sources were variable 
over this time span with typically smooth variations in flux 
with quasi-periods of months to years. VAST will produce light curves
and detect this sort of variability for more 
than 10$^4$ sources over the whole sky. The {\it Fermi} gamma-ray satellite
observes the whole sky every 3 hours and has detected blazar variability on
timescales as short as 6 hours \citep{abdo10a}. 
Simultaneous measurements of
the gamma-ray and radio variability in blazars will provide us with a 
powerful tool for understanding the physics of the central engines in AGN.

Beyond single black holes in AGN, a number of binary supermassive
black holes (SMBHs) are known or suspected
\citep[e.g.,][]{komossa06,rtz+06,smith10,burkespolaor10}. The evidence for these
varies, but it is clear that with the mergers of galaxies the central
SMBHs could also merge.  The act of inspiralling may lead to radio
transients accessible to \tel.  This is because the SMBHs will likely
move through a static magnetic field provided by an accretion disk,
and this should drive electromagnetic jets
\citep{palenzuela10,lyutikov11,moesta12}.
The detailed emission processes are unclear, but it is plausible that
there could be a radio signature accessible to \tel\ before
\citep{oshaughnessy11} or during \citep{kaplan11} the
merger. Numerous other electromagnetic mechanisms have also been
proposed; see \citet{schnittman11} for a detailed discussion.
Combining the radio observations with observations at other wavelengths
or even with gravitational waves would significantly improve our
understanding of galaxy mergers \citep[e.g.,][]{sesana11}.

\subsubsection{Tidal Disruption Events}
Tidal disruption events (TDEs) occur when a star wanders too close to
a black hole at the centre of a galaxy or star cluster.  The black
hole tidal forces tear the star apart, but the debris continue on
ballistic orbits around the star and are accreted as they approach
periastron on the following orbit.  The resulting sudden increase in
the black hole accretion rate may be directly detected at optical and
X-ray frequencies. After the initial flare up the luminosity decays as
$t^{-5/3}$ \citep{rees90}.  Several candidates of such TDEs have been
detected serendipitously through all-sky X-ray and ultraviolet surveys
\citep{grupe99,komossa99a,komossa99b,esquej07,esquej08,gezari06,gezari08,gezari09,lin11,gezari12,cenko12,saxton12}.

Crucially, black hole accretion is often accompanied by relativistic
jets, with a mechanical power often comparable to the radiative
luminosity. The jet can be a source of beamed radio emission and hard
X-rays in a way similar to a blazar.  To a hard X-ray telescope such
as the Burst Alert Telescope (BAT) on the {\em Swift} it will
initially resemble a GRB, but it will be distinguished by its much
longer duration.  {\em Swift}~J1644$+$57 is the first observed example
of such a TDE candidate, at redshift $z = 0.35$.  It was discovered as
a luminous X-ray transient and was observed across the electromagnetic
spectrum, enabling multi-band analysis and modeling of the event
\citep{bloom11,levan11,zauderer11,berger12,wiersema12}.  The mass of
the central black hole responsible for this event was derived to be
$\lesssim 10^7 M_\odot$ and the ejecta Lorentz factor $\gtrsim 10$.
More recently another TDE candidate has also been detected by Swift in
hard X-rays, {\em Swift}~J2058.4+0516 \citep{cenko11}.

The main difference between AGN and TDE jets is that the former propagate
to Mpc distances, and create characteristic cocoon/lobe structures
where their kinetic power is dissipated. By contrast, TDE jets
are impulsive ejections that produce a forward shock/reverse shock
structure as they interact with the ISM \citep{sari95}. As there is
no pre-existing cavity around the previously quiescent black hole, TDE jets
have to propagate through a much denser environment, and decelerate
to sub-relativistic speed at distances $\lesssim 1$~pc. Most of their kinetic
power is expected to be dissipated in the reverse shock \citep{giannios11}.
Typical TDE radio flares from the disruption of
a solar-mass star by a $10^7$ solar mass black hole are expected to peak
in brightness after $\sim 1$~year, reaching a flux density of $\sim 2$~mJy at 1.4~GHz,
for a distance of 1 Gpc corresponding to $z=0.2$ (not including the effect
of relativistic boosting for jets oriented along our line of sight).
At peak brightness, the expected radio spectrum is inverted (self-absorbed)
with $F_\nu \sim \nu^{-2}$ below $\sim 2$~GHz, and $F_\nu \sim \nu^{-1/3}$ above that
\citep{giannios11}.

Radio afterglows of TDEs would be a powerful tool to understand
jet physics and source geometry, through measurement of the
flux evolution and polarization.
The expected degree of polarization for these events
is low if they are collimated, similar to GRB afterglows. 
Recent infra-red and radio polarization measurements have been found to be consistent
with theoretical expectations \citep{wiersema12,metzger12a}. 

Radio detections of TDE jets hold the key to other unsolved astrophysical
problems \citep[e.g.][]{krolik12}. It is well known that not all accreting black holes
have jets, but it is not clear why. Jets are formed only at certain ranges of accretion
rate, but other parameters such as the black hole spin may also be important.
Finding what fraction and what type of X-ray detected TDEs have radio jets will help address
this question, and will shed light on the evolution of nuclear black holes in galaxies.
Furthermore, the timescale for the formation of a radio-bright TDE jet
will help us understand how quickly a collimated magnetic field is generated
inside the transient accretion disk.

The predicted brightness of TDEs at gigahertz frequencies suggests that
VAST should be able to detect about a dozen events per year.
A systematic survey of the radio sky would
be able to constrain the rate of TDEs and reduce
the collimation uncertainty significantly.
In particular, the discovery of an entirely new population of orphan TDEs
would allow us to place solid constraints on the occurrence rate of this
phenomenon and therefore test the theoretical prediction that we should see
$\sim$10$^{-4}$ yr$^{-1}$ per galaxy \citep{rees90}.

\subsubsection{Accreting Neutron Stars, Black Holes and Microquasars}
Exploring the connection between accretion disks and jets in X-ray binaries (XRBs)
requires correlated radio and X-ray observations \citep{fender06}.
The VAST survey will reveal and subsequently measure the
radio emission from a large number of XRBs and allow us to search for new
radio outbursts. This, coupled with the recent availability of large
field of view instruments like {\it Swift}-BAT and {\it Fermi}-GBM, 
and with the early results (from 2013) of eROSITA's all sky survey, 
will provide a large sample of XRBs with active jets, and will allow us to do statistical 
studies of their time variability properties and X-ray/radio correlation.

VAST will yield information on the
population of XRBs showing quasi-steady state radio jets,
which are mildly relativistic, and are associated with 
non-thermal, radiatively-inefficient accretion states
\citep{fender04,mcclintock06}. For these persistent jets, 
fundamental properties such baryonic content, energy budget, 
emission mechanism, and the role played by the spin of the compact object 
are not well constrained \citep{markoff05,narayan96,yuan05},
and regular radio monitoring will allow us to investigate them.
In addition, we will determine the duty cycle of the radio-loud phases, 
and test whether radio state transitions (changes in the kinetic power) 
are always associated with X-ray transitions (changes in the radiative 
power). We will also investigate what physical mechanism  
causes major radio flares and the ejection of highly relativistic, 
optically-thin bullets, often associated to the state transition 
from the hard/non-thermal to soft/thermal state.

In addition to stellar mass and supermassive black holes, it is predicted
that VAST will detect jets from intermediate mass black holes (IMBHs) with
masses between $\sim$100 -- 100,000 M$_\odot$. IMBHs may provide a pathway
for the formation of supermassive black holes \citep[e.g.][]{ebisuzaki01}
and also have important connotations for other areas of astrophysics 
\citep[e.g.][]{fornasa08,wang10}, but until
recently their existence was highly disputed. The discovery of the object
known as HLX-1 located in the galaxy ESO 243-49, with an X-ray luminosity
$\sim$1000 times greater than the Eddington limit for a 10 M$_\odot$ black
hole, currently provides very strong evidence for the existence of IMBHs
\citep{farrell09}, with independent studies constraining the mass to between  9,000 M$_\odot$ $<$ M$_{BH}$ $<$ 90,000 M$_\odot$ \citep{davis11,servillat11,godet12,webb12}.
Recently, transient radio emission was detected from HLX-1
following a transition from the low/hard to high/soft spectral states,
consistent with the ejection of relativistic jets from an IMBH \citep{webb12}.

Objects such as as HLX1 may be very rare (only one known within 100 Mpc), 
but there are hundreds of ultraluminous X-ray sources 
(ULXs; see \citealt{feng11} for a recent review) in the local Universe 
with luminosities exceeding the classical Eddington limit of stellar-mass 
black holes. 
Most of them are likely to be explained through other mechanisms such as
anisotropic emission and/or super-Eddington accretion, 
but we cannot rule out that a few of them may be IMBHs. In addition, 
there may be hundreds of low-state or quiescent IMBHs lurking in the core 
of globular clusters, in the nuclei of dwarf galaxies, or inside recently merged galaxies that may 
occasionally become active.
The detection of jet emission from these objects would allow us to infer the
black hole mass, as the radio and X-ray luminosities are correlated as a
function of mass \citep[e.g.][]{fender06}.
Steady jets would be undetectable for
all but the closest of these objects. However, transient radio flares have
been observed from XRBs associated with jet ejection events with radio
luminosities a factor of 10 -- 100 times brighter than the non-flaring radio
emission \citep[e.g.][]{kording06}. Such flares 
should be detectable by VAST out to distances of
$\sim$5 -- 15 Mpc from 1,000 M$\odot$ black holes, and out to $\sim$23 -- 72
Mpc from 10,000 M$\odot$ black holes (Farrell, in preparation). Thus we will 
be able to survey the local Universe for flaring IMBHs, placing
constraints on the number density and therefore testing their validity as a
pathway for supermassive black hole formation.
Conversely, if ULXs are super-Eddington accretors, detection of 
radio emission will prove that even at those extreme accretion rates, 
a substantial fraction of accretion power can come out as kinetic power, 
and will improve our understanding of black hole feedback 
at high accretion rates.

\subsubsection{Neutron Star Mergers}\label{s_nsm}
The merger of two compact objects (neutron stars) is expected to
generate strong gravitational waves (\S\ref{s_gw}) but they will also
reveal themselves through electromagnetic radiation. The relativistic,
beamed outflows detected from short-hard GRBs are thought to originate
from such mergers \citep[e.g][]{fong12}. VAST is not expected to
fortuitously observe such beamed events. However, merger simulations
also show that substantial mass can be ejected from these systems
quasi-isotropically and at sub- or trans-relativistic velocities, and
with energies of 10$^{49}-10^{50}$ erg \citep{piran12}.

Consider a mildly relativistic blast wave ($\beta_0\sim 0.8$) moving
out with an energy $\sim 10^{49}\,$erg into the surrounding
medium. 
The shock-wave sweeps the surrounding medium and heats it to
relativistic temperatures by converting bulk kinetic energy of the
incoming material into thermal energy of the shocked material. The
electrons in the shock-heated plasma are accelerated in the post-shock
magnetic field to radiate synchrotron radiation.  This would
eventually be optically thin in the gigahertz regime, peaking on a
timescale of about a couple of months when the blast wave starts to
decelerate. The optimum waveband for detecting these transients, which
are also supposed to be prime sources of gravitational waves, is 
around gigahertz frequencies
\citep{nakar11}. A merger within 300~Mpc would shine as a radio 
transient of about 1~mJy brightness at $\sim$1~GHz (Kamble \& Kaplan in preparation)
which would be detectable to the VAST and, 
potentially, also to Advanced-LIGO.
Both the peak flux density and the time to peak are
uncertain since we lack accurate estimates of the ejecta
velocity and energy, and we expect the circumburst density to vary
from 1 cm$^{-3}$ to 10$^{-4}$ cm$^{-3}$.

Using the `best-bet' neutron star merger detection rates predicted for
Advanced LIGO \citep{abadie10}, we estimate that up to about half a
dozen of these transients should be detectable by ASKAP at any given
time, although there are significant uncertainties involved.
\citet{piran12} independently estimate ASKAP's sensitivity to these
events and obtain similar results. These are comparable to the rate
estimated from observations of short GRBs which have been proposed to
originate in similar mergers \citep{coward12}.  \citet{metzger12b}
emphasise the challenges related to uncertain GW event localisations,
which are significant. As discussed in \S\ref{s_gw}, the large
instantaneous sky coverage of ASKAP may be a significant advantage in
this situation.

\subsubsection{Flaring and Pulsing Neutron Stars}
Of the known neutron stars, the vast majority are pulsars that
have been detected through their pulsed radio emission.  VAST will not
be sensitive to radio pulsations on timescales of $\lesssim 1$~second,
but will instead probe neutron star populations that
are not selected in pulsar surveys.  We will be sensitive to
intermittent or deeply nulling radio pulsars
as well as radio pulsars with large variations in their average pulsed flux.

Magnetars will be detectable through their rare flaring events such as
the giant flare and expanding radio nebula produced by SGR~1806$-$20
\citep{gaensler05,cameron05}.
In addition, some magnetars are known to turn on as radio sources.
For example, XTE~J1810$-$197
was initially detected as a transient radio 
source \citep{halpern05} before its radio pulsations were 
identified \citep{camilo06}.  We may also
be able to detect bright intermittent pulses from the enigmatic
rotating radio transients \citep[RRATs;][]{mclaughlin06}.
VAST will thus reveal a broader census of the Galactic neutron star population
than a standard pulsation search, with subsequent implications for the
overall supernova event rate in our Galaxy.

The radio source FIRST J1023+0038 presents an interesting case study.
Initially it was classified as a Galactic cataclysmic variable (CV) \citep{bond02} because of
its optical spectrum and variability, but its radio flaring behavior
appeared to be anomalous.  The classification was called into question
because of the changes in the spectrum over time
\citep[e.g.,][]{thorstensen05} and it was eventually re-identified as a
transitional object where a recycled radio pulsar had turned on after
a low mass X-ray binary phase \citep{archibald09,archibald10}. The
actual transition event had an extremely short duration, with the
accretion disk apparently dissipating between 2001
December and 2002 May. The radio pulsar was probably absent in
1998 (upper limits of 1.8 and 3.4~mJy at 1.4 GHz) but detected in a survey
in mid-2007 with a flux density $\sim$14 mJy at 1.6~GHz.

\subsubsection{Classical novae}
Novae occur in binaries (CVs) that consist of
a white dwarf that is accreting matter from its companion star.  A
nova is a thermonuclear explosion in the layer of accreted matter on
the surface of a white dwarf.  A shell of matter is ejected during
the explosion and this shell might produce radio emission.  The radio
emission can persist for a few years after the explosion, and in the
case of FH Ser it reached $\sim$10~mJy at 2.7~GHz about one year after
the outburst.  In the late stages when the ejecta is optically thin,
radio emission is generated by the entire ejecta and can therefore be
used to estimate the total ejecta mass (see e.g., \citealt{hje1990a}
for a review).

\subsubsection{Flare stars, pre-main sequence stars, and
  rotation-driven stellar activity}

There are several types of stars that display large flares that can be observed
in the radio band.  The most common are cool dwarf stars of spectral type M
or later including brown dwarfs.
The flares have typical flux densities above $\sim 1$~mJy,
but there is also quiescent emission at the sub-mJy level between the
flares (e.g. \citealt{pestalozzi00}).
For these types of systems it has been argued that radio observations are
the best way to obtain information on the stellar magnetic field strength
\citep{berger06}.
The number of flaring events that are potentially detectable is unclear, but
\citet{berger06} obtained a detection rate of $\sim 10\%$ for objects of later
spectral class than M7.  They also find that the radio to X-ray luminosity
ratio increases over at least three orders of magnitude for these objects compared
to earlier spectral classes.
VAST will undertake a census of such stars in the local (10\,pc) neighbourhood.

In AM Her-like systems (cataclysmic variables in which the white
dwarf has a surface magnetic field of $\sim 10^3 - 10^4$\,T) the M dwarfs are 
forced to co-rotate with the binary, which has an orbital period of a few hours.
The dwarf star then becomes more active; flares in AM Her itself have 
been observed to be as bright as 10~mJy at 5~GHz \citep{dulk83}.

Another group of stars that can produce radio flares are the RS CVn-systems
\citep{gunn94}.
These are close detached binaries consisting of two main sequence or
evolved stars of spectral type F or later, that are tidally locked.
The rapid spin of the stars leads to strong magnetic activity, which is
manifested as large spots on the stars and also as flaring activity
that can be observed both in X-rays and at radio wavelengths.

Flare-like outbursts are also observed in pre-main sequence stars, but
it is not clear whether the physical mechanism is really that of a
stellar flare or related instead to the accretion flow onto the star.
Such outbursts have been observed in both naked T Tauri stars, which
do not show signs of extended accretion discs, and in systems with an
infrared excess, which indicates the presence of a disc
\citep{osten09}.

A magnetic field can convert the rotational energy of a star into an
outflow and electromagnetic radiation.  The most well-known examples
of this are the classical radio pulsars, but a similar phenomenon can
arise in less extreme stars.  Radio observations have detected two
fully circularly polarised radio pulses per spin period from the Ap
star CU Vir, which is also spinning down in a way similar to a pulsar
\citep[e.g.][]{tri2008a,trigilio11,lo12}.

Another example of activity driven by rotation is the white
dwarf in the cataclysmic variable AE Aqr.  The spin period of the
white dwarf is 33\,s, so rapid that the white dwarf does not accrete
the matter that streams over from the companion K dwarf.  Instead, the
matter is ejected from the system.  AE Aqr can be observed over a wide
range of frequencies from X-rays to radio, and pulsed emission is
observed at optical and X-rays, but it has not yet been detected at
radio wavelengths \citep{dej1994a,boo1987a,bas1988a}.

\section{Exploration of the Unknown}\label{s_blind}
The most interesting sources detected by VAST may be those we currently
know nothing about.
A sensitive
wide-area blind survey is ideal for detecting such unknown source
classes, but one of the challenges in designing the VAST survey is ensuring
that the survey parameters do not unduly bias us against
unknown source classes that we would otherwise be sensitive to.
For example, daily monitoring of a field for 5 minutes per day makes it
less likely that we will detect an object that brightens on hour-long 
timescales and fades within several hours.
\begin{figure}[t!]
\begin{center}
\includegraphics[width=\columnwidth]{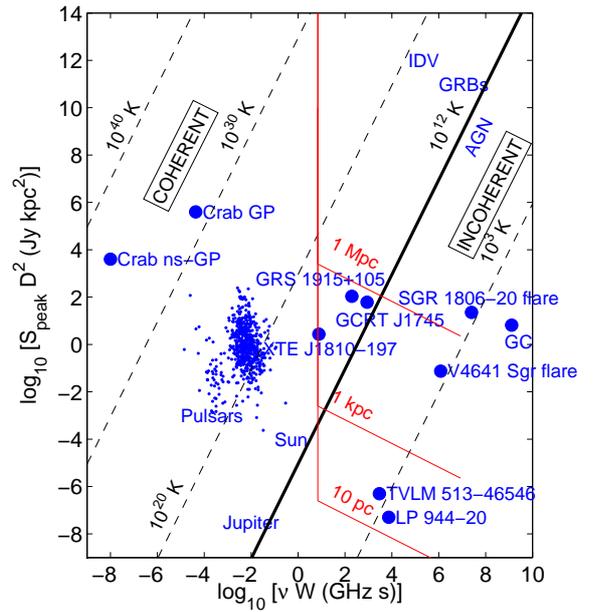}
\caption{\changes{The parameter space for radio transients, adapted from
  \citet{cordes04}. A quantity equivalent to absolute luminosity
  (observed flux density $S$ multiplied by the square of the distance
  $D^2$) is plotted against the dimensionless product of the emission
  frequency $\nu$ and the transient duration or pulse width $W$.  In
  the Rayleigh-Jeans approximation, these quantities are directly
  proportional and related to the brightness temperature, as indicated
  by the diagonal lines, with $T=10^{12}$~K marking the maximum
  brightness temperature of incoherent processes. Examples of known
  radio transient sources are indicated, including giant pulses and
  ``nanogiant'' pulses from the Crab pulsar \citep{hankins03}, radio
  pulses from XTE~J1810$-$197 \citep{camilo06} and other pulsars from
  the Australia Telescope National Facility pulsar catalogue
  \citep{manchester05}; the microquasar GRS~1915+105
  \citep{mirabel94}; radio flares from V4641~Sgr
  \citep{hjellming2000}, the brown dwarf LP~944$-$20 \citep{berger01},
  and the magnetar SGR~1806$-$20 \citep{gaensler05,cameron05}; the
  Galactic centre radio transient J1745$-$3009 \citep{hyman05}; pulses
  from the ultracool dwarf TVLM~513$-$46546 \citep{hallinan07}; as
  well as radio emission from the Sun and Jupiter.  Red lines indicate
  the expected sensitivity of ASKAP to sources at distances of 10~pc,
  1~kpc and 1~Mpc.}}
\label{f_phase}
\end{center}
\end{figure}

Large-scale blind surveys for radio transients have been performed by
a number of groups \citep[e.g.,][]{amy89,katz03,matsumura07} over the
last few decades.  These
surveys have been severely limited in their sensitivity, sky coverage,
and cadence.  In spite of this, they have revealed that the
radio sky contains many transient objects, the identities of
some of which remain mysterious.

For example, \citet{hyman05} discovered a bursting transient towards the Galactic
 Centre which lasted for only a few minutes with a flux density in excess of 2~Jy 
at 330~MHz. Subsequent observations showed that the bursts repeated multiple times
with a period of $\sim77$~minutes and a burst length of $\sim10$~minutes.
The identification of this source remains unclear \citep[e.g.][]{kaplan08,roy10},
although similarities with bursts from ultracool dwarfs \citep{hallinan07}
are intruiging. Another example is the well-known ‘Lorimer burst’, discovered in an 
archival search of Parkes pulsar data. \citet{lorimer07} discovered a single 
millisecond pulse of extragalactic origin with the astonishing peak flux density 
of 30~Jy at 1.4~GHz (also see \citealt{keane11} for a second example).
Although the nature of this source has been questioned, and 
recent work suggests the burst may be terrestrial \citep{burkespolaor11}, 
these new discoveries illustrate that the parameter space for radio variability
has not been fully explored yet (Figure~\ref{f_phase}).

In this section we summarise existing blind and archival surveys for 
radio transients, focusing on `slow' transients with 
variability on timescales of seconds or greater.

\subsection{Transient Rates from Blind Surveys}\label{s_bsum}
\citet{bower07} presented the results from an archival survey of 
VLA data. Since the publication of this work there have been 
a number of blind radio transient surveys conducted on both archival data, 
and from the first datasets produced by next generation radio telescopes. 
Most of these have had the dual aim of (a) searching for individual objects 
of interest and trying to identify them; and (b) characterising the population 
statistics of radio transients in order to prepare for future surveys.
Although the reported transient rates from \citet{bower07} have recently
been revised, after a reanalysis of the data by \citet{frail12}, the work
established a framework for comparison of radio transient surveys.

\begin{figure*}[t!]
\centering
\includegraphics[width=\textwidth]{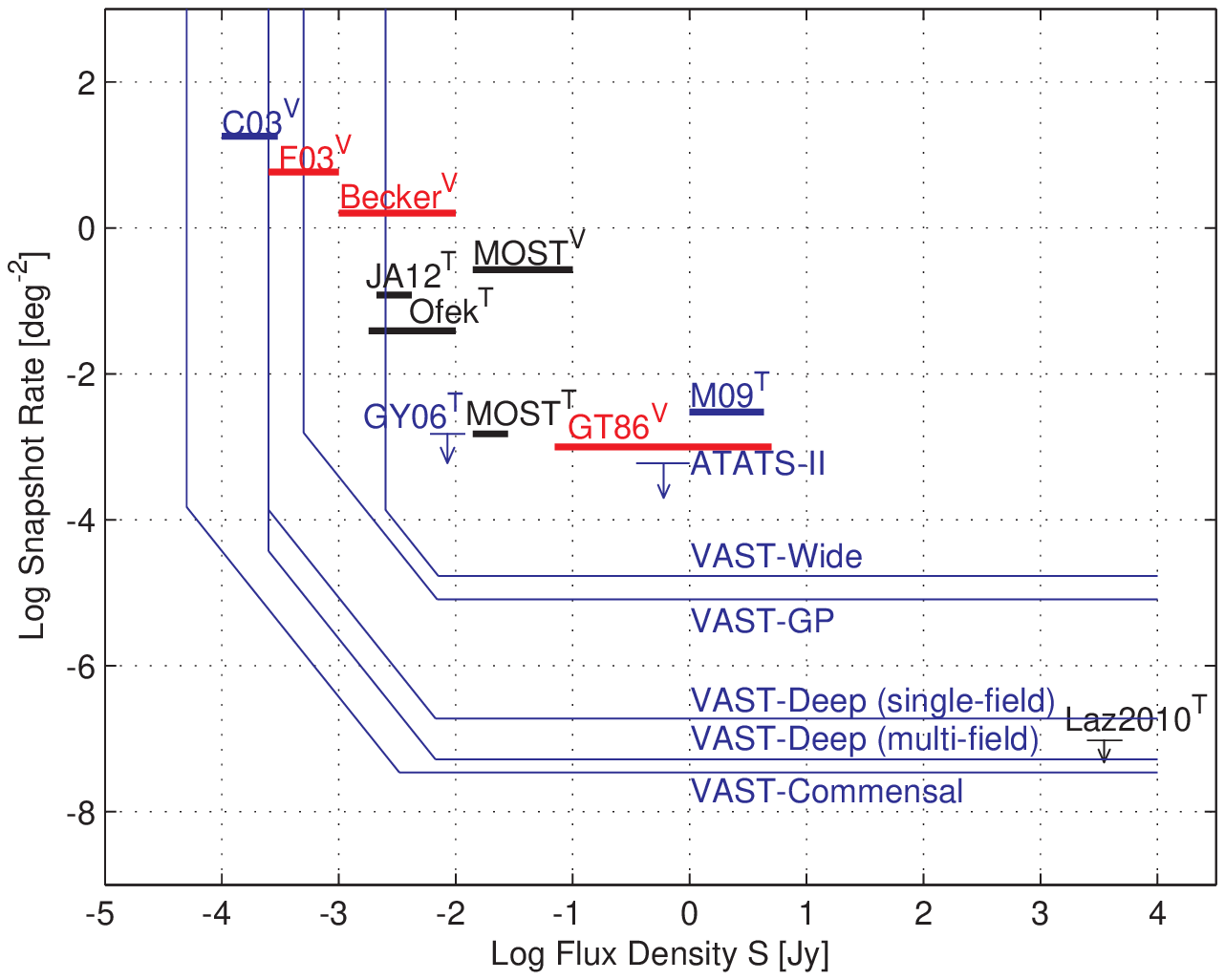}
\caption{\changes{Log 2-epoch snapshot transient rate (deg$^{-2}$) 
against log of the flux density (Jy) for surveys that report detections of transient and 
variables (thick lines), we also include a selection of surveys that report upper limits (thin lines with arrows)  --- see Table \ref{t_blind} for survey acronyms. The surveys are coloured according to frequency: black $<$ 1 GHz, blue = 1 $-$4.8 GHz and red $>$ 4.8 GHz (see Table \ref{t_blind} for details). No corrections have been made for spectral index effects.} 
Surveys labelled superscript `V' denote 
detections of highly variable radio sources; those with superscript `T' denote 
detections of transient type sources.
The VAST survey predictions are indicated and organised by sub-survey. A description of
each survey is given in Section~\ref{s_survey}. In each case 
the vertical segment denotes the RMS that can be achieved per observation. 
The horizontal segment indicates the upper limit that would be set if no transients or 
variables were detected in the entire survey. The joining line (between the horizontal 
and vertical) indicates the upper limits that could be placed by combining data from 
multiple five second snapshots up to the total available integration time per 
observation (see Table \ref{t_blind}). See Section~\ref{s_bsum} for further discussion.}\label{f_snapshot}
\end{figure*}

\citet{bower07} proposed a metric for comparing the results from blind 
surveys for radio transients: the two-epoch equivalent snapshot rate, 
which is equivalent to the transient surface density.
Figure~\ref{f_snapshot} shows an updated 
version of their plot, incorporating a subset of the surveys listed in Table \ref{t_blind}
and discussed in the rest of this section. 
For surveys that made no detection, an upper limit is calculated using a Poisson
distribution:
\begin{equation}\label{e_pn}
P(n) = e^{-\rho A}
\end{equation}
where $P(n)$ is the confidence level (i.e. $P(n) = 0.05$ for 95\%
confidence), $\rho$ is the areal density of transients in a two-epoch survey, 
equivalent to the snapshot transient rate, and $A$ is the equivalent
solid angle, calculated by multiplying the number of images $N_{T}$ by
the field of view $\Omega$.  We can rearrange Equation~\ref{e_pn} in
terms of $\rho$ for the purpose of placing theoretical constraints on
planned surveys:
\begin{equation}\label{e_rho1}
\rho = -\frac{\ln P(n)}{\Omega \times N_{T}} .
\end{equation}

We can improve Equation~\ref{e_rho1} by taking into account
not only the total number of observations in a given survey $N_{T}$, but the intra-observation 
cadence $\tau$, out of total observing time per observation $T$. Doing this results in:
\begin{equation}\label{e_rho2}
\rho = -\frac{\ln P(n)}{\Omega \times N_{T} \times (T/\tau)} .
\end{equation}
For example, on the smallest timescale, ASKAP images will be produced every 
$\tau=5$ seconds; the total observing time per observation (in one scenario) could 
be $T=1$ hour and the field could be observed daily for two years ($N_{T} = 730$).  
A five second ASKAP image will have an RMS of 1.4~mJy~beam$^{-1}$. If 
subsequent data or images are stacked together (either via the uv or image plane) the 
RMS will improve to:
\begin{equation}\label{e_sigma}
\sigma = \frac{1.4 \mbox{ mJy beam}^{-1}}{\sqrt{N_{T} \times (T/\tau)}} 
\end{equation}
Equations~\ref{e_rho2} and \ref{e_sigma} have been used to generate survey predictions 
from the VAST survey parameters described in Section~\ref{s_survey}, and these predictions 
are shown in Figure~\ref{f_snapshot}.  
We also include the predicted rates of a subset of known radio transients 
\citep[see][for futher discussion]{frail12}.
One limitation of the plot in Figure \ref{f_snapshot} is that it collapses
a number of dimensions into a single two-dimensional plot. For example, it 
does not represent the cadence of the surveys, or any trends in transient
detection rate with observing frequency.

Surveys typically fall into one of two categories: many repeated 
observations of a single field \citep[e.g.][]{carilli03,bower07}, or a large 
area, few-epoch surveys \citep[e.g.][]{levinson02,galyam06}.
The two-epoch equivalent snapshot rate is a valid metric for quantising the
effectiveness of a survey, provided the 
the timescale of transient behaviour is less than the cadence of the
observations but longer than a single integration. In this case, a survey with many epochs is equivalent to
a larger area survey with two epochs. Further survey metrics have been
developed to account for observation cadence and source evolution 
timescales as well \citep[e.g.,][]{ska97}.

\begin{sidewaystable*}
\centering
\caption{Summary of snapshot rates for transient and variables radio sources reported in the literature. The results are organised according to upper limits based on non-detections (top section); transient detections (middle section); and detections of highly variable radio sources (bottom section).}
\begin{tabular}{cccccccccc}
\hline Survey/Reference & S$_{Min}$ & S$_{Max}$ & Rate & Timescale & Frequency & Epochs \\
& (mJy) & (mJy) & (deg$^{-2}$) & & (GHz) & (N)\\
\hline
\cite{bower07} & $>$0.09 & --- & $<$6 & 1 year & 4.8 \& 8.4 & 17 \\
\cite{bower07} \& \cite{frail12}$^{\dag}$ & 0.2 & --- & $<$3 & 2 months & 4.8 \& 8.4 & 96 \\
\cite{bower07} \& \cite{frail12}$^{\dag}$ & $>$0.37 & --- & $<6\times10^{-1}$ & 20 mins - 7 days & 4.8 \& 8.4 & 944 \\
PiGSS-I/\cite{bower10}(A)$^{\star}$ & $>$1 & --- & $<$1 & 1 month & 3.1 & 75 \\
PiGSS-II/\cite{bower11b}(B)$^{\star}$ & $>$ 5 & --- & $<$0.18 & 1 month & 3.1 & 78 \\
FIRST-NVSS/\cite{galyam06} & $>$6 & --- & $<$1.5$\times10^{-3}$ & days to months & 1.4 & 2 \\
\citet{bell11} & $>$8 (8$\sigma)$ & --- & $<$0.032 & 4.3 - 45.3 days & 1.4, 4.8 \& 8.4 & 5037 \\
PiGSS-I/\cite{bower10}(A)$^{\star}$ & $>$10 & --- & $<$0.3 & 1 month & 3.1 & 75 \\
PiGSS-II/\cite{bower11b}(B)$^{\star}$ & $>$ 15 & --- & $<$0.025 & 1 day & 3.1 & 78 \\
ATATS - I/\cite{croft10} & $>$40 & --- & $<$0.004 & 81 days - $\sim$ 15 years & 1.4 & 12\\
\cite{bower11}(A)$^{\star}$ & $>$70 & --- & $<$3$\times10^{-3}$ & 1 day & 1.4 & 1852 \\
ATATS - II/\cite{croft11} & $>$350 & --- & $<$6$\times10^{-4}$ & minutes to days & 1.4 & 12 \\
\cite{bower11}(B)$^{\star}$ & $>$3000 & --- & $>$9$\times10^{-4}$ & 1 day & 1.4 & 1852 \\
\cite{lazio10} & $>$2.5$\times10^{6}$(5$\sigma$) & --- & $<$9.5$\times10^{-8}$ & 5 mins & 0.0738& $\sim$1272 \\
\hline
JA12$^T$\cite{jaeger12} & 2.1 & & $<$0.12 & 1 day to 3 months & 0.325 & 6 \\
Ofek$^{T}$/\citet{ofek11} & 2.4 & --- & 0.039 & 1 day--2 years & 4.9 & 16 \\
MOST$^{T}$/\citet{bannister11a}$^{\ddag}$ & 14(5$\sigma$) & 6500 & $1.5\times10^{-3}$ & days--years & 0.843 & 3011 \\
M09$^{T}$/\cite{matsumura09} & 1000 & 4300 & 3$\times10^{-3}$ & $\sim$ 1 day & 1.4 & --- \\
\hline
C03$^{V}$/\cite{carilli03} & $>$0.1 & --- & $<$18 & 19 days and 17 months & 1.4 & 5 \\
F03$^{V}$/\cite{frail03} & $>$0.25 & --- & 5.8 & $\sim$ 1 day & 5 \& 8.5 & --- \\
Becker$^{V}$/\cite{becker10} & $>$1 & --- & 1.6 & $\sim$ 15 years & 4.8 & 3 \\
GT86$^{V}$/\cite{gregory86} & 4.6 & --- & 1$\times 10^{-3}$ & 1 day - 5 years & 4.9 & 16 \\
MOST$^{V}$/\citet{bannister11a} & $>$14 & --- & 0.268 & days--years & 0.843 & 3011 \\
\hline
\end{tabular}
\label{t_blind}
\begin{flushleft}
\changes{$^{\dag}$Note that although \citet{bower07} report 10 detections, the reanalysis by \citet{frail12} results in only 4 (uncertain) detections. The snapshot rate calculated by \citet{frail12} assumes no detections, and we adopt this as the more 
conservative position for our analysis.}\\
$^{\star}$\citet{bower10}, \citet{bower11b} and \citet{bower11} each state two different rates depending on flux density and timescale; we quote these separately as (A) and (B).  \\
$^{\ddag}$See also \citet{bannister11b}.
\end{flushleft}
\end{sidewaystable*}

\subsection{Archival Surveys}
The first large radio archival survey of variability and transient behaviour was 
conducted by \citet{levinson02} who covered a large fraction of the sky with two epochs
by comparing the NVSS \citep{condon98} and FIRST \citep{becker97} surveys, both at 1.4~GHz. 
Various corrections were made to account for the different
footprint of each survey and their differing resolutions
(45\arcsec\ for NVSS compared to 5\arcsec\ for FIRST). 
Nine possible radio transients were reported by \citet{levinson02},
but follow-up of these sources with the VLA by \citet{galyam06} ruled out five of these
as false candidates and two as non-variable sources.
The effective survey area was 5990~deg$^2$, hence with two detections the overall transient
snapshot rate was $\rho = 1.5\times10^{-3}$~deg$^{-2}$.

Another large area survey was conducted by \citet{bannister11a}, using archival data
from the Molonglo Observatory Synthesis Telescope \citep{mills81} which operates at 843~MHz.
This dataset had the advantage of a common resolution and comparable sensitivity in each
epoch, but like the NVSS-FIRST comparison had a range of cadences between days and
years for each field. \citet{bannister11a,bannister11b} detected 2 transient sources from a survey
with a two-epoch equivalent area of 2776~deg$^2$, giving an overall snapshot rate of 
$\rho = 1.5\times10^{-3}$~deg$^{-2}$ for sources above 14~mJy.

An alternative approach has been to use archival calibrator fields which have many 
repeated observations but a relatively small field of view. \citet{bell11} analysed
5037 observations of VLA calibrator fields at three frequencies (1.4, 4.8 and 8.4~GHz).
The observations ran from 1984 to 2008, with typical gaps between observations of between 4.3 and
45.3 days. No radio transients were detected and they therefore placed an upper limit 
of $\rho < 0.032$~deg$^{-2}$ for sources brighter than 8.0~mJy between 1.4 and 8.4~GHz.

Likewise \citet{bower11} detected no transients in their analysis of archival observations of 3C~286
from the VLA. Their dataset spanned 23 years and included 1852 epochs at 1.4 GHz.
They derive upper limits of $\rho < 3\times10^{-3}$~deg$^{-2}$ at 70~mJy for timescales of $\sim1$ day and $\rho < 9\times10^{-4}$~deg$^{-2}$ at 3~Jy for timescales of $\sim1$~minute.
These results, combined with the \citet{croft10} results discussed in Section~\ref{s_ghz} below, 
currently provide the best limits on the number of slow, bright radio transients.

\subsection{Gigahertz Surveys}\label{s_ghz}
There have been a number of blind imaging surveys specifically designed to search for 
transient and highly variable sources. 
Until recently these have generally been small area surveys with many epochs, due to
the limited field of view of the telescopes used.
Early surveys were motivated by searching for radio counterparts to GRBs and 
supernovae.  For example, \citet{frail94} observed 15~epochs of a field centred on the location 
of GRB~940301 with the Dominion Radio Astrophysical Observatory Synthesis Telescope
at 1.4~GHz. These observations were taken between 3 days and 99 days after the initial 
event and made no detections, which resulted in an upper limit of 3.5~mJy on any time-variable
radio sources.

\citet{carilli03} observed the Lockman Hole region to a sensitivity limit of $100\mu$Jy
at 1.4~GHz
with five epochs over a period of 17 months. They found that 2\% of the sources within their sample were highly
variable (defined as $\Delta S \ge 50\%$) and hence derived a rate of $\rho = 18$~deg$^{-2}$. Also in search 
of highly variable radio sources, \citet{becker10} surveyed the Galactic plane and found 
$\rho = 1.6$~deg$^{-2}$ above a sensitivity 1~mJy at 4.8 GHz.
At low Galactic latitude and sampling a multitude of timescales (days to years) \citet{ofek11} used the VLA 
to study transient and variables in a field covering 2.66~deg$^{2}$ to a sensitivity limit of $\sim100 \mu$Jy
at 5~GHz. 
One transient candidate was reported and a thorough discussion of the variable radio sources within the field was given.   

 \citet{matsumura09} summarise the detections of nine candidate transient
sources from the Nasu 1.4~GHz wide-field survey 
\citep[see][for full details]{kuniyoshi07,matsumura07,niinuma07,kida08,niinuma09}. 
The sources were detected in a drift scan mode, and have flux densities 
greater than 1~Jy with typical timescales of minutes to days. (Note,
however, the questions raised about these results by \citealt{croft10,croft11} and the
discussion about the resulting transient event rates by \citealt{ofek11}.)

As briefly discussed earlier in Section~\ref{s_bsum}, \citet{bower07} reported the detection of ten transient radio sources in archival VLA data at 4.8 and 8.4 GHz. \citet{frail12} recently reanalysed this dataset and reported that more than half of these transients were either caused by rare data reduction artifacts, or that the detections had a lower S/N after re-reduction.    
The \cite{bower07} study initially predicted a rate of events over the whole sky of $\rho \sim$ 1.5~deg$^{-2}$, which is diminished to potentially an order of magnitude less, depending on how the lower S/N level transients are interpreted. 
\changes{For Figure~\ref{f_snapshot} we have adopted the conservative snapshot rate calculated by \citet{frail12}, which assumes no
detections. See \citet{frail12} for further discussion.}

The Allen Telescope Array \citep[ATA;][]{welch09} is the first of the next generation telescopes to have been
used for large area transient surveys. The Pi GHz Sky Survey (PiGSS) survey was carried out on the ATA at 3.1~GHz 
and covered  10\,000~deg$^{2}$ of the sky with a minimum of two epochs for each field with an RMS 
sensitivity of $\sim$1~mJy \citep{bower10}.

The first data release (PiGSS-I) is a 10~deg$^2$ region in the Bo\"{o}tes constellation, with 75 daily observations
over a 4~month period. \citet{bower10} identify one object present in PiGSS that is not in archival
radio catalogues (NVSS and WENSS). From this they set an upper limit on the snapshot rate of transients
of $\rho < 1$ deg$^{-2}$ at 1~mJy sensitivity at 3.1~GHz. The second data release (PiGSS-II; \citealt{bower11b}) focused on 
studying the daily and monthly transient and variable nature of the field. 
The upper limits of $\rho < 0.025$ deg$^{-2}$ at 15~mJy for one day timescales and $\rho < 0.18$ deg$^{-2}$ 
at 5~mJy for one month timescales were placed on the rates of transient events at 3.1~GHz.

The other major ATA survey is the The Allen Telescope Array Twenty-Centimeter Survey 
\citep[ATATS;][]{croft10}, a 690~deg$^{2}$ survey with 12 epochs of each field.
\citet{croft10} presented results from comparing a single combined ATATS image with the NVSS. They
did not detect any transients, placing an upper limit of $\rho<0.004$~deg$^{-2}$ on the transient
rate for sources with flux greater than 40~mJy at 1.4~GHz. \citet{croft11} extended this analysis presenting results for an individual epoch-to-epoch 
comparison for the ATATS dataset. They derived a limit of $\rho < 6\times10^{-4}$~deg$^{-2}$
for transients brighter than 350~mJy at 1.4~GHz.

\subsection{Low Frequency Surveys}
Since many significant discoveries of radio transients have been made at low frequencies,
the low frequency SKA pathfinders have enormous discovery potential.

The largest blind survey for slow transients at low frequencies was conducted by 
\citet{lazio10} with the Long Wavelength Demonstrator Array (LWDA).
The LWDA is a 16 dipole phased array with all-sky imaging capabilities, operating at 73.8~MHz.
A total of 106~hours of data 
(with a typical integration time of 5~minutes) was searched for
transients. No detections were made above a flux density limit of 500~Jy. 
\citet{lazio10} quote an upper limit on the rate of $1 \times 10^{-2}$ yr$^{-1}$deg$^{-1}$
which converts to a rate of $\rho < 9.5\times10^{-8}$~deg$^{-2}$ assuming 5 minute integrations.  

In comparison, one of the deepest low frequency studies to date has been conducted by \citet{jaeger12}. 
They used six VLA observations centred on the SWIRE Lockman Field at 325~MHz to study sources brighter than 
0.2~mJy within a field of view of 6.5 deg$^{2}$. The timescales of observation ranged from 1 day to three months. 
One transient was reported 
and the authors argue that it is most likely coherent emission from a stellar flare.  Using this one detection, 
\citet{jaeger12} set an upper limit on 
the snapshot rate of $\rho <0.12$ deg$^{-2}$ for sources brighter than 2.1~mJy at 325~MHz. 

Between the two comprehensive low frequency transient surveys discussed above 
(\citealt{lazio10} and \citealt{jaeger12}) lies a large amount of parameter space both in frequency and time 
to explore. This will be made possible by two new instruments: the Murchison Widefield Array \citep[MWA;][]{lonsdale09,tingay12} and the Low Frequency Array 
\citep[LOFAR;][]{heald11}, both currently being
constructed and commissioned.

\section{Survey Strategy}\label{s_survey}
The detection and characterisation of radio transients requires both 
the sampling of new regions of parameter space, and the ability to monitor 
and follow up known or newly detected sources.  As such, a successful 
transient instrument requires a large field-of-view, rapid survey 
capabilities, high sensitivity and dynamic range, as well as 
sufficient angular resolution for unambiguous source identification 
and follow-up. \tel\ will fulfill these criteria with its ability to 
achieve sub-millijansky sensitivity across the entire visible sky in a 
single day of observing.

Sources that vary on rapid timescales are 
necessarily compact. The 10\arcsec\ resolution of \tel\ will permit source 
localisation to $\sim$arcsecond positional accuracy, enabling 
follow-up observations at other wave-bands.

The key to the efficient detection and monitoring of transient and
variable sources is repeated observation of each field to sample
a full range of cadences corresponding to the natural timescales of
variability in the underlying source population.
In addition to observing commensally with other ASKAP programs, we
will run a campaign of complementary surveys designed to achieve our
science goals.  The VAST surveys build on the standard ASKAP imaging
pipeline and thus do not include other possible modes of operation
such as ``Fly's Eye'' monitoring \citep[e.g.,][]{siemion12}.

\subsection{VAST Surveys}\label{s_sspecs}
In this section we outline our strategy for each of the VAST surveys given
in Table~\ref{t_survey}.
We have based this survey strategy on an estimate of $\sim 8500$ hours of
observing time over the first five years of ASKAP operations, in addition
to continuous commensal observing with other ASKAP projects.
The survey parameters are summarised in Table~\ref{t_survey}.
All of the dedicated (non-commensal) VAST surveys will be conducted with an observing frequency range of 
1130--1430~MHz.

\subsubsection{VAST Wide}
To detect rare events such as GRBs and SNe, and to map out the gas density distribution 
of the ISM across the whole sky, a large survey area is required.
To fully sample the ESE light curves, a regular (preferably daily) cadence is
desirable. 
Two years of data are required to discriminate between an annual cycle
and intrinsic variability for IDV sources.

Hence our planned survey will cover $10\,000$ square degrees (comprised of 400 pointings, each of 
30 square degrees). 
Integrating for 40 seconds per pointing
will give an expected RMS sensitivity of 0.5~mJy~beam$^{-1}$, with a total observing time (including overheads)
of 6 hours per day. This survey will be repeated daily
for 2 years, with a total time of 4380 hours.
Figure~\ref{f_snapshot} shows that it will be a substantial improvement over
existing blind gigahertz surveys.

\subsubsection{VAST Deep}
In our Deep survey we plan to observe $10\,000$ square degrees, with 400 pointings
each observed for 1 hour. This would achieve an expected RMS sensitivity of 
50~$\mu$Jy~beam$^{-1}$. Each of the fields will be observed 
by the EMU collaboration (see Section~\ref{s_commensal}) for their continuum survey \citep{norris11}.
While both the Deep and
the Wide surveys observe the same amount of sky, they
are complementary in cadence and sensitivity. 

One field, coinciding with a field observed by the EMU collaboration,
will be repeated daily for one year (Deep Single field). The other fields
will be repeated 8 times at irregularly spaced intervals (Deep Multi-field).
This is the optimal survey to detect
as yet unknown source classes \citep[see][]{bower07} and detect GRBs
and SNe to larger distances. 

\subsubsection{VAST Galactic}
A third component of VAST is a survey of 750 square degrees of the Galactic plane.
This is a total of 30 pointings, plus an additional 3 pointings to cover the Large
and Small Magellanic Clouds. 
Each pointing will be observed for 16 minutes, achieving an expected RMS 
sensitivity of 0.1~mJy~beam$^{-1}$.
The Galactic plane survey will require 9 hours of observing
in a single day, repeated weekly for 1 year and supplemented by additional
epochs at longer intervals.

\subsection{Commensal Observing}\label{s_commensal}
In addition to the planned surveys, we will operate our transient
detection pipeline commensally on all other ASKAP observations.
Two projects, EMU \citep[The Evolutionary Map of the Universe;][]{norris11}\footnote{See http://www.askap.org/emu.} and WALLABY 
(The ASKAP HI All-Sky Survey; Koribalski, et al., in preparation)\footnote{See http://www.askap.org/wallaby.}, are particularly relevant as they 
will conduct extensive (possibly commensal) wide-field surveys:
\begin{description}
\item[EMU:] a deep ($10~\mu$Jy~beam$^{-1}$ RMS) survey of the entire Southern Hemisphere, to a declination of
$\delta = +30^\circ$.
\item[WALLABY:] a survey of neutral Hydrogen covering two thirds of the 
sky to a redshift of $z=0.5$.
\end{description}
These projects plan to operate in `point and shoot' mode,
covering some region of the sky with significant integration time ($\sim12$~hours) per pointing 
and no revisits. This provides an excellent opportunity for detection and monitoring of transient and
variable behaviour on timescales of less than a day.

\subsection{VAST Predictions}
In Figure~\ref{f_snapshot} we show the predictions parameterised via the method described 
in Section~\ref{s_blind}, using the VAST survey parameters given in Table \ref{t_blind}. 
Note, the horizontal lines show the upper limits that would be placed if no transient 
or variable sources were detected. The vertical lines show the RMS that would be expected 
for the final integration time per image (see RMS sensitivity in 
Table~\ref{t_blind}). For the commensal survey the best RMS is found after six hours 
(not 12), assuming that two back-to-back six hour images are needed to 
search for transients, i.e., that there 
are no repeat observations.  It can be seen from Figure~\ref{f_snapshot} that 
the VAST surveys make orders of magnitude improvements on previous surveys.  

VAST-Wide will use 4320 hours of dedicated observing time, which is the greatest of all the 
VAST sub-surveys (excluding commensal observing). It will observe 10\,000 deg$^{2}$ for 40 seconds 
every day for two years, possibly longer. Observing for 40 seconds with ASKAP yields an RMS 
of $\sim$0.5~mJy~beam$^{-1}$ and produces eight five-second images per observation. This is represented 
on Figure \ref{f_snapshot} and it casts VAST-Wide as less competitive than, for example, 
VAST-Deep (both multi-field and single field). The limits shown in Figure \ref{f_snapshot} for VAST-Wide 
however do not include upper limits on the rate and RMS calculations via combining data from multiple days observing. 

For example, all the 40 second images taken daily for an entire week could be combined together 
and compared and searched for transients with the combined images from
the next week, and so on. If technically feasible, such an approach will
significantly increase the competitiveness of the VAST-Wide survey.
Figure~\ref{f_snapshot} also does not convey how many \emph{known} radio sources will be monitored 
in each case. VAST-Wide will monitor all known sources in the Southern hemisphere above 2.5 mJy 
($5\sigma$) every day for two years, whereas VAST-Deep (one field) will monitor only a subset 
(30 deg$^{2}$) to a greater sensitivity of 250$\mu$Jy ($5\sigma$). We can approximate the ratio 
of sources that will be observed with VAST-Deep ($N_{Deep}$) to VAST-Wide ($N_{Wide}$) using the 
following expression (assuming a Euclidean source population):  
\begin{equation}
\frac{N_{Deep}}{N_{Wide}} =  \frac{A_{Deep}}{A_{Wide}} \times \left (\frac{\sigma_{Deep}}{\sigma_{Wide}}\right )^{-1.5} \simeq 10\%
\end{equation}
\noindent where in this example $A_{Deep}$ is the survey area of
VAST-Deep (single field, 30 deg$^{2}$); $A_{Wide}$ the area of
VAST-Wide (10,000 deg$^{2}$). $\sigma_{Deep}$ and $\sigma_{Wide}$
represent the survey sensitivities respectively. VAST-Wide will
therefore monitor large numbers of bright transients and variables,
such as IDVs and AGN. VAST-Deep will be better optimised for studying
the faint end of the transient and variable population.

\section{Data Processing}\label{s_software}

The success of VAST hinges on our ability to detect transient sources 
on the timescales on which images are produced by ASKAP, at the same time as 
monitoring known variable sources that are in our field. 
Firstly, how to effectively deal with the large volumes of data, and secondly, how
to find the `needles in a haystack' --- the most interesting transient sources.
As discussed in Section~\ref{s_blind}, current blind surveys at 1.4~GHz suggest that less
than $0.5\%$ of sources 
observed by VAST will be transient or highly variable \citep{bannister11a}.

\begin{figure}[t!]
\includegraphics[width=8cm]{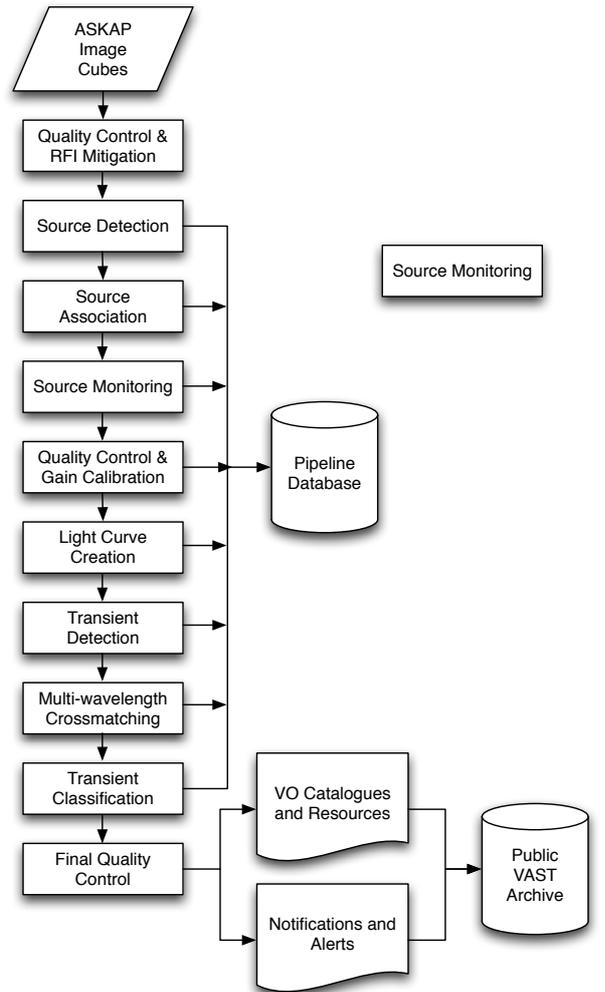}
\caption{\changes{Illustration of the VAST pipeline functionality. A description of each stage 
is given in Table~\ref{tab:pipeline_func}.}\label{fig:pipeline}}
\end{figure}

The operation of the ASKAP data pipeline is described in the Science Processing 
Document \citep{cornwell11}. The following sections address some of the issues most
relevant to VAST, and discuss the design and operation of the VAST transient detection pipeline.
The overall functionality of the VAST pipeline 
is illustrated in Figure~\ref{fig:pipeline} and described in Table~\ref{tab:pipeline_func}.
A discussion of the pipeline design is also provided by \citet{banyer12}.

\subsection{Data Rates}
ASKAP is designed to be a real-time telescope, with most of the data processing 
done in online mode. Data from the phased array feed elements in each antenna will be 
sent to the beam former to create primary beams. The primary beams from each antenna 
are sent to the correlator to create visibility data. These data are then sent to the 
central processor for processing into science-ready data products: images and spectral 
cubes. The data rate from the correlator to the central processor is expected to be
2.5~GB/s, generating about 200~TB of raw data every 24 hours.

The central processor will operate three data processing pipelines in parallel: one pipeline
will produce data cubes designed for spectral line analysis, with high frequency resolution;
a second will produce data cubes designed for continuum surveys, with full polarisation and higher
angular resolution, but reduced spectral resolution; the third pipeline is the transient
detection pipeline which will produce data cubes with lower frequency resolution,
but operating on 5-second timescales (see Table~\ref{t_vastim}). Each 5-second image cube 
will be up to 16~GB (depending on the number of frequency channels 
generated and stored), which means that the transients pipeline will
process around 12~TB of data an hour. Retaining some coarse frequency
resolution will enable post-correlation rejection of radio frequency
interference (see section~\ref{sec:rfi} below) and also provide some
minimal spectral information for detected transient sources.

\begin{table}[t!]
\begin{center}
\caption{VAST Image Products}\label{t_vastim}
\begin{tabular}{lc}
\hline
Polarisations & 1--4 \\
Frequency channels & 30 \\
Field of view & 7.5 deg$^2$ \\
Angular resolution & 10\arcsec \\
Integration time & 5~seconds \\
Image size (pixels) & $4096\times4096$ \\
Image cube size & up to 16~GB \\
\hline
\end{tabular}
\end{center}
\end{table}

\subsection{Transient Detection Approach}
Source detection is central to all upcoming radio transient surveys, and also 
those in other wave-bands such as the Large Synoptic Survey Telescope transients survey.
In the image domain, transient source detection has typically followed one of 
two approaches: image subtraction or catalogue-based detection. 

Supernova search projects and other optical astronomy projects have typically
used the image subtraction approach \citep[for example,][]{riess04,alcock00}. 
In its simplest form, this involves subtracting the 
current image from some higher sensitivity master image.
In practice, some corrections are required, for example each image can
be convolved with a kernel to account for coordinate shifts and distortions \citep{alard98,alard00}.

In radio images, the incomplete u-v coverage means that image subtraction often
results in artefacts that can then be mistakenly identified as transient sources.
A more robust approach is to firstly extract and measure sources in each image, and
then search for transients by catalogue cross-matching and lightcurve analysis.
The approach taken in the VAST transient detection pipeline is based on techniques 
described by \citet{bannister11a}, developed for analysing the Molonglo Observatory Synthesis
Telescope archive. This is also similar to the approaches taken in the Allen Telescope Array
transients projects \citep{croft10,croft11} and the LOFAR transient detection pipeline \citep{swinbank07}.

We are also investigating the technique of transient detection in the visibility domain 
\citep{trott11,law12}.
These approaches have the advantage of dealing with well-behaved noise
properties of visibilities and avoiding artefacts caused by the
imaging and deconvolution processes.
\citet{trott11} use statistical decision theory to derive detection limits of visibilities
using models of the source and calibration effects of interferometers.
\citet{law12} present a computationally-simple,
calibration-independent algorithm for transient detection based on the
bispectrum, an interferometric closure quantity. By subtracting
visibilities in time, the bispectrum can be used to quantify the
significance of transients throughout the field of view.
Visibility-based algorithms have great potential in the study of
transients, because they take advantage of \emph{a priori} information
to reduce computational burden. This makes real-time processing
easier, which may be a critical limitation of the large,
computationally-demanding surveys described here. More work is needed
to understand how these algorithms work on a large scale and whether
they can be implemented for real-time analysis.
\begin{table*}[t!]
\begin{center}
\caption{Functionality of the VAST pipeline components, as illustrated in Figure~\ref{fig:pipeline}.\label{tab:pipeline_func}}
\begin{tabular}{p{1.5in}p{3.5in}}
\hline
\noalign{\smallskip}
Function & Description\\
\noalign{\smallskip}
\hline
\noalign{\smallskip}
\changes{Quality control \& RFI mitigation} & 
\changes{Initial quality control on the image cubes output from the ASKAP pipeline. Poor quality images will be flagged or ignored.
Possible RFI mitigation strategies are discussed in Section~\ref{sec:rfi}}.\\
\noalign{\smallskip}
Source detection &
Blind source finding is performed on every image as discussed in Section~\ref{s_sf}. 
This produces a list of detected sources with position, flux, size and local RMS.\\
\noalign{\smallskip}
Source association &
Each detected source is associated with one from the master list by finding the master source at the same position as the detection. If no master source is found at that position then a new master source is added.\\
\noalign{\smallskip}
Source monitoring &
Master sources known to be in the field that were not detected by the source finder are measured by attempting to fit a Gaussian at the position where the source is expected to be. If no source can be fit then an upper limit on the flux at that position is taken instead.\\
\noalign{\smallskip}
\changes{Quality control \& gain calibration} & 
\changes{The image is assessed for quality by comparing the measured parameters of the detected sources to a list of 
well-characterised sources using the techniques described by \citet{bannister11a}.}\\
\noalign{\smallskip}
Light curve creation &
All measurements for each source are collated into a radio light curve --- a collection of flux measurements for all epochs where the source was observed.\\
\noalign{\smallskip}
Transient detection & 
The updated light curves are analysed to detect transient or variable behaviour. These sources will go on list for monitoring, and cross-matching with external catalogues. \\
\noalign{\smallskip}
\changes{Multi-wavelength cross-matching} & 
\changes{Sources of interest are cross-matched with external catalogues to provide additional information on the likely causes of variability, and to collect extra information for classification and identification.} \\
\noalign{\smallskip}
\changes{Transient classification} & 
\changes{Sources that exhibit transient or variable behaviour are classified using a machine-learning algorithm. Possible approaches are discussed in Section~\ref{s_class}.} \\
\noalign{\smallskip}
\changes{Quality control \& final data products} &
\changes{After a final quality control stage, sources that appear genuine with a high degree of confidence will be released to the
community via VOEvent notifications as discussed in Section~\ref{s_alerts}. These alerts, along with science quality catalogues of transient and variable sources will be made public through the VAST archive.} \\
\noalign{\smallskip}
\hline
\end{tabular}
\end{center}
\end{table*}

\subsection{Rejection of Radio Frequency Interference \label{sec:rfi}}
One of the most significant challenges for all radio telescopes is the impact of 
radio frequency interference (RFI) from man-made sources. Although the ASKAP site 
was chosen for its radio-quiet properties, signals from satellites used for communications, 
navigation, remote sensing and the military pervade all areas of the Earth. 
Similarly aircraft communication signals are present even in very remote areas. 
In addition, self-generated RFI is a challenge as faster digital equipment is used in the signal 
pathways. There is no universal method of RFI mitigation that is satisfactory for all types of 
observing techniques and RFI signals \citep{fridman01,briggs05}.
For the data rates expected with VAST, an automated approach with a combination of techniques will be essential.

RFI mitigation techniques range in sophistication from simple blanking of corrupted data 
to adaptive cancellation based on coherent detection. Simple blanking works well for strong short 
duration RFI, but is completely ineffective in removing low-level time-variable RFI that could mimic 
transient sources. Blanking removes signal as well as interference and reduces the sensitivity achieved 
in a given time. There is a growing literature on more sophisticated ideas for RFI elimination. Some of
the techniques that we are exploring for VAST include: 
\begin{enumerate}
\item Post-detection flagging based on recognition of data that are affected by interference. 
This has been the traditional approach in radio interferometry and is appropriate for moderately 
strong interference, but it does come with a loss of sensitivity. 

\item Adaptive cancellation techniques such as those described in \citet{mitchell05,briggs05,kesteven10}.
These require a separate reference antenna or beam on the sky to sample the RFI.  An algorithm is used that 
adapts filter characteristics to match the RFI spectral footprint. This approach requires a high sensitivity 
copy of the interferer. 

\item Spatial filtering techniques that use the location of the known 
interferer or the relative arrival times of the signals to identify and separate 
the RFI signal from the astronomical signal \citep[e.g.][]{kocz10,leshem00,ellingson02,smolders02}.
These are appropriate for stationary or slowly moving RFI sources and require 
less sensitivity to the interfering source than the adaptive cancellation techniques.

\item Adaptive nulling techniques, which employ the capacity of the array to place 
response nulls on the RFI source. These are appropriate for very strong interfering sources 
and for telescopes such as ASKAP with independently steerable beams.

\item Post-correlation filtering techniques use software algorithms to remove RFI from interferometric 
data by subtraction of a weak, persistent RFI signal \citep{athreya09} or by implementation of low-pass 
time and frequency filters to remove variable RFI \citep{offringa12}.
The advantage of these techniques is that they generally have a lower sensitivity requirement 
for detecting the interference. However, radio interferometers may not be able to use this approach 
to mitigate some types of slowly varying RFI signals.  
\end{enumerate}

\subsection{Source Detection}\label{s_sf}
Finding and characterising sources in each of the observed images are
critical first steps for transient detection in the image domain. False
positives in source finding have the ability to confound the detection
of transients, reducing the confidence with which a transient can be
detected. This has been identified as a problem in previous surveys, with
source-finding errors contributing to the false detection rate
\citep[see, for example,][]{bannister11a, croft11, frail12}.

The source finding packages that are currently available (for example the 
Miriad package sfind \citep{hopkins02} and the AIPS package
\textsc{vsad} \citep{condon98}) are all able 
to achieve a high degree of completeness and reliability, typically better
than $95\%$ on standard radio continuum images. These packages have been
extensively used in large-scale radio surveys. However, even a small
number of missed real sources, or false positives, can cause problems in 
the identification
of variable and transient sources. 

In recent surveys the best source
finding results have been obtained by using the above mentioned source
finding packages in conjunction with a pipeline of other programs 
\citep[see, for example,][]{mauch03,murphy07,norris06}. The
refinement of the source finding pipeline is often done after all
observations have been completed and the idiosyncrasies of the data
are well understood. A lot of time and effort is spent normalising the
input images, and filtering the resulting catalogues. In many cases a
subset of the final source catalogue is validated `by eye' to ensure
that the brighter sources are correctly represented. This approach
produces high quality catalogues and is well suited to single epoch
surveys, as is evidenced by the high level of completeness and
reliability that these surveys are able to obtain.

For VAST, the volume of data will require a
fully automated source finding process. The inability to archive every
image that is observed for VAST will require that the source finding
is done in real-time, and without the possibility of reprocessing. The
large number of sources that are expected for a single day's
observation means that even with completeness and reliability of
$99\%$, a source finding algorithm will be finding up to $\sim 10,000$ false
detections and missing up to $\sim 10,000$ real sources each day. These
false detections and missed sources far outweigh the number of
true transient and variable sources that we expect to detect.

\begin{figure}[t!]
\fragnames
\centering
\includegraphics[width=\linewidth,bb=50 190 550 605]{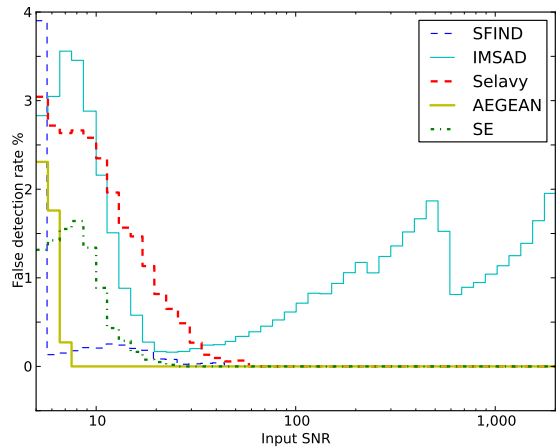}
\caption{ The false detection rate (FDR) for each of the source
finding algorithms \citep[from][]{hancock12}. No falsely detected
sources are expected above an SNR of 5 for the area of sky
simulated. The non-zero rate of false detections is due to poor source
characterisation.}
\label{f_fdr}
\end{figure}
\begin{figure}[t!]
\fragnames
\centering
\includegraphics[width=\linewidth,bb=50 190 550 605]{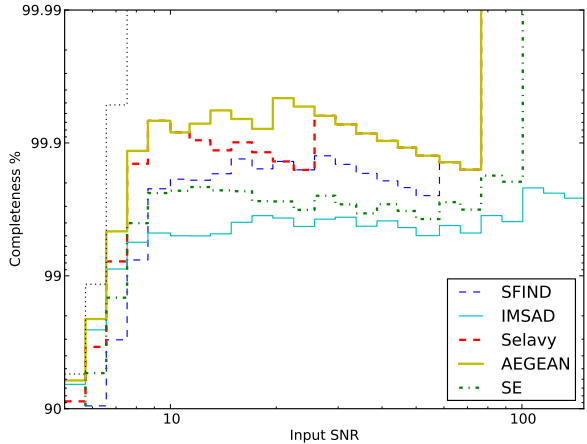}
\caption{ The completeness of each catalogue \citep[from][]{hancock12}
as compared to the input catalogue. The coloured curves represent the
completeness of the named source finders. The black dotted curve
represents the expected completeness of an ideal source finder. The
plateau in completeness above and SNR of 10 is due to sources that are
poorly characterised.}
\label{f_completeness}
\end{figure}

As part of the VAST design study process, \citet{hancock12} 
have undertaken a detailed analysis of some of the most widely used
source finding packages in astronomy, including the newly
developed Selavy \citep{whiting12} and Aegean \citep{hancock12}
algorithms. For compact sources, we find that the task of source {\em
finding} (distinguishing between source and background pixels) is a
solved problem that the various programs all achieve equally
well. However, the task of source {\em characterisation} (measuring
the various parameters of the found sources) is only accomplished well
in the case of single isolated sources. For diffuse sources, the task
of source finding is equally well achieved, however the task of source
characterisation is not well defined. \citet{hales12} address some of
the issues of extended source measurement. However, since we expect transient
sources to be compact, this is not a significant issue for VAST.

Figures~\ref{f_fdr} and \ref{f_completeness} show the false detection
rate (FDR) and completeness of each of the source finding algorithms.
The FDR and completeness were measured by injecting a known population
of sources into a simulated image and processing this image with each
of the different source finding algorithms. A comparison between the
input and output catalogues resulted in: a set of sources common to
both (real detections), a set of sources in the input but not the
output (non-detections), and a set of sources that were in the output
but not the input (false detections). The set of falsely detected
sources consisted entirely of sources within an island of pixels that
contains multiple components. These sources were so poorly
characterised that they were not able to be linked to a source within
the input catalogue. Sources that are poorly characterised also reduce
the catalogue completeness. In Figures~\ref{f_fdr} and
\ref{f_completeness}, the main reason that the source finding
algorithms have non-zero FDR and less than ideal completeness is due
to poor source {\em characterisation}, rather than poor source {\em
finding}. In order to better characterise islands of pixels that
contain more than a single component the Aegean program analyses the
curvature of an image. As can be seen in Figures~\ref{f_fdr} and
\ref{f_completeness}, Aegean is able to characterise islands of pixels
that contain multiple components better than any of the other source
finding approaches, typically achieving a completeness of over
$99.9\%$.  Further details of the source finding analysis and the
operation of the Aegean source finding algorithm are presented by
\citet{hancock12}.

\subsection{Classification of Transients}\label{s_class}
Given the large number of sources that we will likely detect with VAST, prompt classification will be
critical for early identification and multi-wavelength follow-up.
This is a challenging problem due to the sparsity and heterogeneity of the available data.  
There are two sets of information that can be used for this task: the light curve itself, and the 
existing archival information about that source or location on the sky.

Light-curve classification has been successfully used to classify variable stars at optical 
wavelengths \citep[e.g.][]{richards11,richards12}. In order to explore the feasibility of light-curve 
classification for VAST, we used several approaches.

Firstly, we simulated light curves of five transient source types: SNe, RS CVn flare stars (fStar\_RSCVn), 
M-class flare stars (fStar\_dMe), XRBs, novae; two variable source types: IDVs and ESEs and a 
non-varying background (BG) for comparison (Lo et al. in preparation).
We experimented with light-curve 
classification on the simulated light curves in two scenarios: 
\begin{itemize}
\item {\it archival classification} where the classification 
is performed after a significant amount of light curve data has been collected; and 
\item {\it online classification}, which is classification of partial light curves in real-time 
for the purpose of triggering appropriate follow-up observations. 
\end{itemize}
Secondly, we used light curves from the Palomar-Quest and Catalina 
Real-Time Transient Surveys to experiment further with a range of machine learning methods.
The results of these experiments are discussed in the sections below.

\subsubsection{Archival Classification}\label{s_classradio}
We approached the classification problem using the standard methodology of supervised machine 
learning with 10-fold cross validation. First we extracted a set of features from each light-curve. 
These include: time domain features which are 
simply the flux measurements at each time point ({\it tme}), statistical features ({\it stat}) such 
as the mean, standard deviation and skew, following \citet{richards11}; statistical (stat) features 
built cumulatively ({\it stat-cum}) after observing a fixed number of observations in multiples 
of $k$ ($k$, $2k$, etc.); periodic features derived from the Lomb-Scargle periodogram ({\it lsp}); 
and wavelet features derived from the Haar wavelet ({\it wlet}). We then created two more feature sets that are concatenations of the above.  Feature set {\it all-reps} is the concatenation of {\it stat-cum}, {\it lsp}, and {\it wlet}, and feature set {\it all} concatenates {\it tme}, {\it stat-cum}, {\it lsp}, and {\it wlet}.  Details of all seven feature sets are discussed by \citet{rebbapragada2012}. 

\begin{figure}[t!]
\includegraphics[width=\columnwidth]{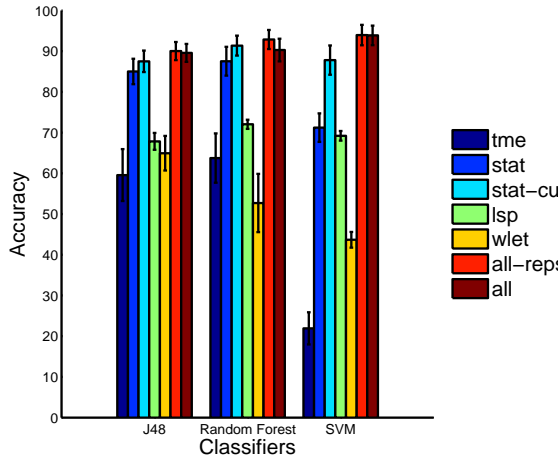}
\caption{Comparison of overall classification performance (accuracy percentage) by feature set ({\it tme}, {\it stat}, {\it stat-cum}, {\it wlet}, {\it lsp}, {\it all-reps}, and {\it all}) and classification algorithm (J48, Random Forest, and SVM), with results grouped by classification algorithm.}
\label{fig:archival_class}
\end{figure}

We used these features as input to a set of machine learning algorithms, which, 
when given a training set of examples, will learn the discriminative features for each 
source type. Using the \textsc{Weka} toolkit 
\citep{weka}, we experimented with three classification algorithms: Support Vector Machines
\citep[SVM;][]{cortes95}, Random Forests \citep{breiman01} and the J48 Decision Tree 
\citep{quinlan86dtrees}. 
Figure \ref{fig:archival_class} shows the classification accuracy in the archival setting. The Random 
Forest is the best performing classifier when used with the combined feature sets, achieving $93\%$ 
accuracy. This suggests archival classification using light curves alone is feasible with a reasonable
degree of accuracy, given the large number of sources. However, the performance on real light curves is likely to be lower since our simulated light curves have ideal Gaussian noise and are aligned to the starting time of the event.

\subsubsection{Online Classification}
The aim of online classification is to identify sources of interest as soon as possible after new
observations arrive. The earlier a classification is made, the more
likely that a follow-up observation can be scheduled, and so online
classification should run continuously as part of the real-time
transient detection pipeline.
We evaluated online 
classification using the first 30 days of simulated light curves. With so few observations, it is not 
possible to extract periodogram or wavelet features, hence we were limited to the statistical feature 
set and the time series values themselves.

\begin{figure}[t!]
\centering
\includegraphics[width=\columnwidth]{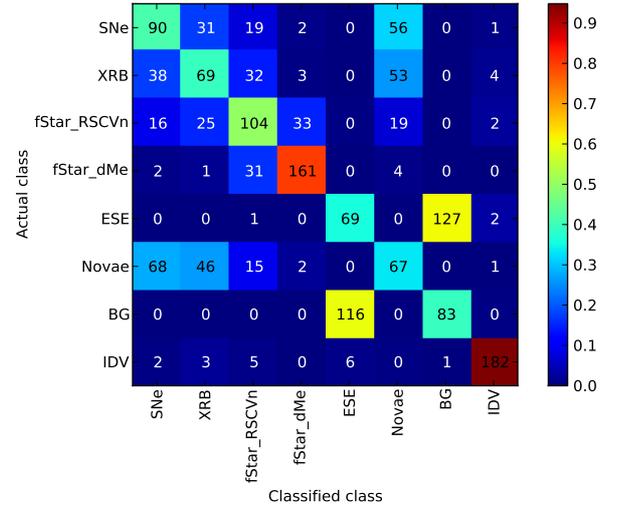}
\caption{Online classification confusion matrix shown as a heat map (red = higher accuracy, blue = lower accuracy). The x-axis is the classified output class and the y-axis is the actual input class. We used 200 sources of each type and the number in each cell represents how these sources are classified. Overall accuracy at 30 days is 50\%.}
\label{fig:online_class}
\end{figure}

Figure \ref{fig:online_class} shows the confusion matrix for our best performing algorithm:
the J48 classifier built with {\it stat} features extracted from 30 days of
observations. The overall 
classification accuracy achieved was $50\%$ across eight source types. This is better than the 
baseline expected from random classification ($12\%$ accuracy) but probably not sufficient to
be useful in scheduling follow-up observations. 
Accuracy was improved by grouping source types with similar light-curve characteristics together 
(grouping SNe and Novae, BG and ESE yielded an accuracy of $70\%$). Finally, accumulating more data led to 
higher accuracy ($\sim80\%$ accuracy with 100 days of observations). 

\subsubsection{Other Machine Learning Approaches}
Considerable insight into the classification problem can be gained from the current efforts on 
classification of transients in optical synoptic sky surveys 
\citep{bloom08,donalek08,mahabal08a,mahabal08b,mahabal10,mahabal11,djorgovski11}.
We experimented with several approaches, using the data from the Palomar-Quest and Catalina 
Real-Time Transient Surveys, as well as selected data sets from the literature.

Similar to the work described in Section~\ref{s_classradio} we
parameterised observed optical light curves as feature vectors, and applied various 
machine learning techniques (e.g., decision trees) to their classification. A set of 
decision trees were trained using the feature vectors for various combinations of classes. 
To reduce the dimensionality of the input space, we applied a forward feature selection 
strategy that consisted of selecting a subset of features from the training set that best 
predict the test data by sequentially selecting features until there is no improvement in 
prediction.  Each tree was built using the Gini diversity index as criterion for choosing 
the split; the splitting stopped when there is no further gain that can be made.  To avoid 
over-fitting we used a 10-fold cross validation approach.
The decision trees were pruned in order 
to choose the simplest one within one standard error of the minimum.  When tested on a 
data set consisting of light curves of blazars, cataclysmic variables, and RR Lyrae stars, 
we achieved completeness in the range of $\sim$ 83\% - 97\%, and contamination in the range 
of $\sim$ 4\% to 13\% (Donalek et al., in prep.). In future work we will apply this to the
classification of radio light curves.

Another approach we experimented with used two-dimensional distributions of magnitude changes for different 
time baselines for all possible epoch pairs in the data set \citep[][Moghaddam et al., in preparation]{djorgovski11}.
These two-dimensional ($\Delta m, \Delta t$) 
histograms can be viewed as probabilistic structure functions for the light curves of different 
types.  Template distributions for different kinds of transients and variables are constructed 
using the reliably classified data with the same survey cadences, S/N, etc.  For any newly 
detected variable or a transient, corresponding ($\Delta m, \Delta t$) histograms are accumulated 
as the new data arrive, and we used a variety of metrics to compute the effective probabilistic 
distances from different templates. Initial testing indicates that we can obtain a high classification
accuracy using this approach. We plan to generalise it to include triplets 
or even higher order sets of data points for multi-dimensional histograms.

One important lesson so far is that existing archival and contextual data will play a critical role 
in classification.  This includes a spatial context (i.e., what is near the observed event on the sky ---
a possible host galaxy, a cluster, a SN remnant, etc.), the multi-wavelength context (has it been detected 
at other wavelengths, what is its broad spectral energy distribution), and the temporal context (what 
was its flux variability or a detection history in the previously obtained data).    
Some of this information can be readily extracted from the archives (e.g., flux measurements from different 
wavelengths, light curves at that location), but some - spatial context in particular - require a human 
judgment, e.g., is the apparent proximity to other objects (galaxies, clusters, etc.) likely to be relevant, 
and if so, what does it imply about the transient?  Human inspection of vast numbers of transients does 
not scale to the massive data streams such as those contemplated here.  We are currently experimenting 
with crowdsourcing approaches to harvesting of the relevant human pattern recognition skills and domain 
expertise, and their translation to machine-processable algorithms.

\subsection{Alerts and Follow-up}\label{s_alerts}
ASKAP policy specifies that all data and results must be released as soon as they have passed
scientific quality control. This is particularly important for transient detections, in which
multi-wavelength follow-up is critical. VAST will generate publicly available data products on
the following timescales:
\begin{description}
\item[Seconds:] Triggers for individual transient events will be released through 
services such as 
VOEventNet\footnote{See http://voeventnet.caltech.edu.} as soon as they are detected. 
These events will have quality flags representing our 
confidence about the detection and will allow immediate follow-up of new discoveries by the 
wider community. 
\item[Minutes:] Triggers with value-added data such as classifications based on 
multi-wavelength archival crossmatching will be released as soon as our source classification 
pipeline has run. 
\item[Days:] Measurements of every source in every field will be entered into our 
lightcurve database in near real-time. This will be available immediately to the VAST 
collaboration. Once quality control has been completed, updates to the database 
will be made publicly available.
\item[Months:] Scientific quality catalogues will be constructed from our 
lightcurve database. The results will be published and data made public as soon as possible 
after analysis is complete.
\end{description}
\changes{The quality control process for alerts will consist of a multi-tiered system in which there
is automatic filtering of a large fraction of false detections followed by manual inspection
of the remaining sources. This approach is taken by several existing transients surveys, for example
in the Palomar Transients Factory \citep{galyam11}. The detailed implementation of this general approach will be
developed and refined during commissioning.}

In addition, all images from VAST will be made available through the standard ASKAP archive.
Where possible the VAST pipeline will use standard Virtual Observatory protocols such as
VOEvents\footnote{See http://www.ivoa.net/Documents/VOEvent.}.

\subsubsection{Very Long Baseline Interferometry}
As noted earlier, radio sources that are highly variable on short timescales 
are necessarily intrinsically compact.
ASKAP's angular resolution of $\sim$10\arcsec
will be high enough to localise variable sources on the sky and
measure variability, but will be orders of magnitude too low to
resolve the evolving structures that give rise to the variations in
integrated flux.  In order to resolve the structure and evolution of
the radio source, much higher angular resolution imaging is required,
provided by Very Long Baseline Interferometry (VLBI).

Alerts from the VAST pipelines will be used to trigger observations
using the Australian Long Baseline Array (LBA), now including
telescopes from Western Australia (ASKAP) to New Zealand (Warkworth),
allowing milliarcsecond-scale imaging.  Due to the very large number
of VAST alerts, targets for VLBI follow-up will have to be chosen
selectively and will need to be restricted to the most interesting or rare
classes of transient and variable sources.  The Australian LBA is very
well suited to this type of follow-up, as evidenced by previous
programs that have taken triggers for transient radio sources from
MOST or the ATCA \citep{tingay95,millerjones12}.

\changes{The current LBA has a sensitivity of $60\mu$Jy~beam$^{-1}$ for a one hour
integration at 2.3 and 8~GHz. This is comparable to the sensitivity of ASKAP at 
1.4~GHz. With ASKAP incorporated as an element of the LBA, the sensitivity at 1.6~GHz 
would be $40\mu$Jy~beam$^{-1}$ for a one hour integration.
Hence a $1−2$ hour observation with the LBA would be able
to detect any unresolved source found by the VAST surveys.}

The recent advent of real-time eVLBI on the LBA means that the VLBI observations
can be rapidly processed and analysed. Thus subsequent observations can be
adapted based on the initial results. This facility will benefit studies of new
transients that will be detected by ASKAP.

\section{Conclusions}
Next generation radio telescopes such as the EVLA, LOFAR, MWA and ASKAP will 
have a substantial impact on our ability to discover and monitor radio 
transient and variable sources. 
The VAST survey will play an important role in the study of a range of phenomena,
from GRB afterglows and supernovae through to cataclysmic variables and radio
flare stars, as well as propagation effects in the interstellar medium.

The three VAST surveys (VAST-Wide, VAST-Deep and VAST-Galactic)
will allow us to do a comprehensive search of radio transient parameter space. 
It is likely that in addition to discovering a large number of known classes of 
objects, we will also discover new types of transient and variable phenomena. 
Our automatic transient detection pipeline will be able to run simultaneously with
other major ASKAP observing projects, meaning that we will also analyse data in 
commensal mode.

All data products produced by VAST will be released to the wider community as soon
as possible after passing quality control standards. Wherever possible this will
be done through the use of Virtual Observatory tools and protocols.

In this paper we have summarised the science goals of VAST, described our current
results in algorithm development for source-finding and transient light curve 
classification, and made predictions about the likely source populations that will
be observed by VAST. 

\section*{Acknowledgements} 
We would like to acknowledge the additional contributors to the optical transient classification work, including 
Ashish Mahabal, Ciro Donalek, Matthew Graham, Baback Moghaddam, Mike Turmon, and a number of Caltech students.
We would also like to thank Peter Ashwell and Andrew Naoum at the University of Sydney for useful discussions.
We acknowledge the contribution of the wider VAST collaboration team members through their participation
in the design study over the last three years.

This research has been supported in part by the Australian Research 
Council (ARC). TM,
BMG, PH, JB and MB acknowledge support through grants FS100100033 and 
DP110102034, and through the Science
Leveraging Fund of the New South Wales Office for Science and Medical
Research. 
SAF is the recipient of an ARC Postdoctoral
Fellowship, DP110102889.
TM, BMG and S. Croft acknowledge funding from the University of Sydney 
International Program Development Fund. 
The Centre for All-sky Astrophysics is an Australian
Research Council Centre of Excellence, funded by grant CE110001020.  

S. Chatterjee acknowledges support from the US National Science
Foundation (NSF) through the award AST-1008213.
DLK was partially supported by NSF awards AST-1008353 and AST-0908884.
SGD acknowledges partial support from the NSF grants AST-0407448, 
AST-0909182 and IIS-1118041, and the National Aeronautics and 
Space Administration (NASA) grant 08-AISR08-0085.
IHS is supported by an NSERC Discovery Grant.
A portion of this research was carried out at the Jet Propulsion Laboratory, 
California Institute of Technology, under a contract with NASA. 
US Government sponsorship acknowledged.

\bibliographystyle{apj}
\bibliography{mn-jour,vast}

\end{document}